%% file: Main.tex
\documentclass[format=acmsmall, review=false, screen=true]{acmart}

\usepackage{booktabs} 
\usepackage{bm}
\usepackage{amsmath}
\usepackage{multirow}
\usepackage{booktabs}
\usepackage{threeparttable}
\usepackage{ulem}

\usepackage{footnote}
\makesavenoteenv{table}

\usepackage{lscape}
\usepackage[ruled]{algorithm2e} 

\SetAlFnt{\small}
\SetAlCapFnt{\small}
\SetAlCapNameFnt{\small}
\SetAlCapHSkip{0pt}
\IncMargin{-\parindent}

\acmJournal{CSUR}
\acmVolume{9}
\acmNumber{4}
\acmArticle{39}
\acmYear{2021}
\acmMonth{5}
\copyrightyear{2021}

\acmDOI{0000001.0000001}

\begin{document}
\title[A Survey on Session-based Recommender Systems]{A Survey on Session-based Recommender Systems}

\author{Shoujin Wang}
\authornote{Corresponding author: Shoujin Wang, shoujinwang@foxmail.com}
\affiliation{%
  \institution{Macquarie University}
  \city{Sydney}
  \country{Australia}}
\email{shoujinwang@foxmail.com}

\author{Longbing Cao}
\affiliation{%
  \institution{University of Technology Sydney}
  \city{Sydney}
  \country{Australia}
}
\email{longbing.cao@uts.edu.au}

\author{Yan Wang, Quan Z. Sheng, Mehmet A. Orgun}
\affiliation{%
 \institution{Macquarie University}
  \city{Sydney}
\country{Australia}}
\email{{yan.wang,      michael.sheng,   mehmet.orgun}@mq.edu.au}

\author{Defu Lian}
\affiliation{%
 \institution{University of Science and Technology of China}
  \city{Hefei}
\country{China}}
\email{liandefu@ustc.edu.cn}

\begin{abstract}
Recommender systems (RSs) have been playing an increasingly important role for informed consumption, services, and decision-making in the overloaded information era and digitized economy. In recent years, session-based recommender systems (SBRSs) have emerged as a new paradigm of RSs. Different from other RSs such as content-based RSs and collaborative filtering-based RSs which usually model long-term yet static user preferences, SBRSs aim to capture short-term but dynamic user preferences to provide more timely and accurate recommendations sensitive to the evolution of their session contexts. Although SBRSs have been intensively studied, neither unified problem statements for SBRSs nor in-depth elaboration of SBRS characteristics and challenges are available. It is also unclear to what extent SBRS challenges have been addressed and what the overall research landscape of SBRSs is. This comprehensive review of SBRSs addresses the above aspects by exploring in depth the SBRS entities (e.g., sessions), behaviours (e.g., users' clicks on items) and their properties (e.g., session length). We propose a general problem statement of SBRSs, summarize the diversified data characteristics and challenges of SBRSs, and define a taxonomy to categorize the representative SBRS research. Finally, we discuss new research opportunities in this exciting and vibrant area.   

\end{abstract}

%
%


%
%



\maketitle

\renewcommand{\shortauthors}{S. Wang et al.}

\input{paperbody.tex}

\end{document}

%% file: paperbody.tex
\section{Introduction}\label{Introduction}

Recommender Systems (RSs) have evolved into a fundamental tool for making more informative, efficient and effective choices and decisions in almost every daily aspect of life, working, business operations, study, entertaining and socialization~\cite{wang2020era}. Their roles have become ever important in the increasingly overloaded age of digital economy where users have to make choices from usually massive and rapidly increasing contents, products and services (which are uniformly called \textit{
items}). A variety of RS research areas have emerged with great success, such as content-based RSs \cite{aggarwal2016content,pazzani2007content}, collaborative filtering based RSs \cite{schafer2007collaborative,ekstrand2011collaborative}, and hybrid RSs \cite{burke2002hybrid} which combine the first two. 

However, those RSs tend to utilize all historical \textit{user-item interactions} (\textit{interactions} for short, referring to the direct or indirect user actions on items, e.g., the list of a user's clicks on items) \cite{cao2015coupling} to learn each user's long-term and static preferences on items. Such a practice is often associated with an underlying assumption that all of the historical interactions of a user are equally important to her current preference. This may not be the reality in the real-world cases and there are two major reasons. First, a user's choice on items not only depends on her long-term historical preference but also depends on her short-term recent preference and the time-sensitive context (e.g., the recently viewed or purchased items). This short-term preference is embedded in the user's most recent interactions \cite{jannach2017session}, which often account for a small proportion of her historical interactions. Second, a user's preference towards items tends to be dynamic rather than static, that is evolving over time. 

To bridge these gaps in RSs, \textit{Session-based Recommender Systems (SBRSs)} have emerged with increasing attention in recent years. Different from the aforementioned RSs, SBRSs learn users' preferences from the \textit{sessions} associated and generated during the consumption process. Each \textit{session} is composed of multiple user-item interactions that happen together in a continuous period of time, e.g., 
a basket of products purchased in one transaction visit, which usually lasts for several minutes to several hours. By taking each session as the basic input unit, an SBRS is able to capture both a user's short-term preference from her recent sessions and the preference dynamics reflecting the change of preferences from one session to another for more accurate and timely recommendations. In this paper, with the term SBRSs, we refer to all RSs that are centered on the session data to recommend the next interaction or the next partial session (i.e., the remaining interactions) in current session, or the next session with multiple interactions (cf. Section \ref{classifi}). This definition covers those narrowly conceived SBRSs in some studies \cite{hidasi2017recurrent,quadrana2018sequence} that recommend the next interaction in the current session only. Also, it covers both session-based and session-aware RSs discussed in~\cite{quadrana2018sequence}.

There are a variety of studies on SBRSs described by different terms in the literature with different settings and assumptions, targeting different application domains. For example, Hidasi et al. \cite{hidasi2015session} built an SBRS on anonymous session data by assuming a strict order over the interactions (e.g., click an item or watch a movie) within each session to predict the next item to click or the next video to watch. Hu et al. \cite{hu1937diversifying} built another SBRS on non-anonymous session data without the order assumption inside sessions to recommend the next item to purchase. Jing et al. \cite{jing2017neural} devised an SBRS on non-anonymous session data with order assumption inside sessions for the next music or movie recommendations.

Although SBRSs are widespread in various domains and many related studies have been conducted, there are many inconsistencies in the area of SBRSs caused by the diverse descriptions, settings, assumptions and application domains. There is not a unified framework that well categorize them and there are no unified problem statements for SBRSs. More importantly, no systematic discussion is available on the unique characteristics of SBRSs including their problem and session data, the research challenges incurred by the characteristics, and the research landscape and gaps in addressing the challenges. There is not a systematic categorization of all the representative and state-of-the-art approaches for SBRSs. These gaps have limited the theoretical development and practical applications of SBRSs. 

To address the above significant aspects and gaps, this paper provides a comprehensive and systematic overview and survey of the field of SBRSs: 
\begin{itemize}
   \item We provide a unified framework to categorize the studies on SBRSs, which can reduce the confusions and inconsistent views in the field of SBRSs.

    \item For the first time, our work proposes a unified problem statement of SBRSs, where an SBRS is built on top of formal concepts: user, item, action, interaction and session. 
    
    \item We provide a comprehensive overview of the unique characteristics of session data as well as the challenges of SBRSs incurred by them. To the best of our knowledge, this is the first such description.
    
    \item A systematic classification and comparison of SBRS approaches are made to provide an overall view on how the challenges have been addressed and what progress has been achieved in the SBRS area.
    
    \item Each class of approaches for SBRSs have been briefly introduced with key technical details to provide an in-depth understanding of the progress achieved for SBRSs.
    
    \item Lastly, open issues and prospects for the SBRS research are discussed. 
    
\end{itemize}

\section{Related Work }
There are a variety of studies on not only SBRSs but also \textit{Sequential Recommender Systems (SRSs)} \cite{wang2019sequential}. SRS is an area closely relevant to but different from SBRSs. Even in the area of SBRSs, there are many different sub-areas, e.g., next-interaction (e.g., purchase an item) recommendations, next-session (e.g., basket) recommendations. As a result, a variety of corresponding specific works described by different terms exist in the literature, including session-based recommendations, next-item/song recommendations, next-basket recommendations, session-based recommender systems, sequential recommender systems, etc. Although quite similar, these works are usually applicable for different scenarios, with different settings and assumptions, belonging to different areas, i.e., SBRSs or SRSs, or some of the above sub-areas. It is not uncommon that these superficially similar but actually different works not only cause confusions between SBRSs and SRSs, but also lead to significant inconsistencies within the area of SBRSs. Below, we first clarify the concepts and differences between SBRSs and SRSs, then provide a framework to categorize the relevant SBRS studies, and clarify the difference between this paper and the related work.


\subsection{SBRSs vs. SRSs}\label{SRS}
\begin{figure*}[!t]
	\centering
	\includegraphics[width=.8\textwidth]{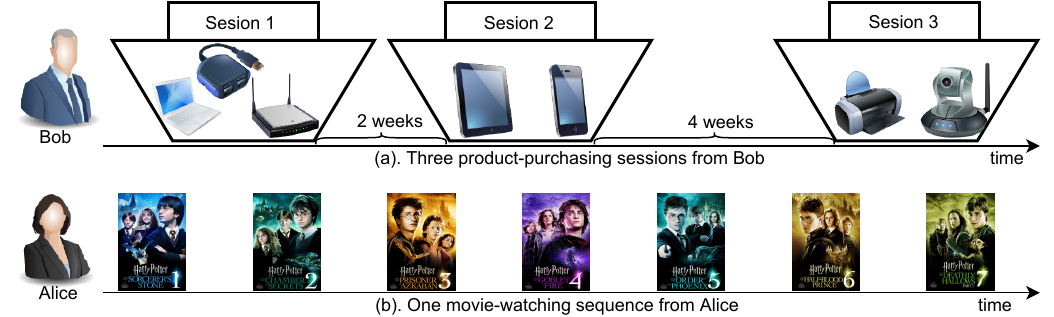}
	\vspace{-0.8em}
	\caption{Session data vs. sequence data}
	\label{fig_sequence}
	\vspace{-1em}
\end{figure*}

SBRSs and SRSs are built on session data and sequence data respectively, while they are often mixed up by some readers. So it is necessary to first clarify the difference between session data and sequence data. A \textit{session} is a list of interactions with a clear \textit{boundary}, while the interactions may be chronologically \textit{ordered} (in ordered sessions) \cite{hidasi2015session,quadrana2017personalizing}, or \textit{unordered} (in unordered sessions)~\cite{hu1937diversifying,wang2017perceiving}. A boundary refers to the starting-ending interaction pair to start and end a specific session in a transaction event. An ordered (unordered) session refers to a session in which the interactions are (not) chronologically ordered. The session data from a given user usually consists of multiple sessions happening at different time and separated by multiple boundaries with non-identical time intervals between sessions (cf. Fig. \ref{fig_sequence} (a)). A \textit{sequence} is a list of historical elements (e.g., item IDs) with clear \textit{order}. The sequence data from a given user often contains a single sequence with only one boundary for it (cf. Fig. \ref{fig_sequence} (b)). In most sequence data, the timestamps are used to sort the elements inside a sequence only while no explicit time intervals are included and considered~\cite{quadrana2018sequence}.    
A boundary usually indicates co-occurrence-based dependencies \cite{cao2015coupling} over the interactions or elements within it. Co-occurrence-based dependencies constitute the foundation of SBRSs, especially for those built on unordered session data. Order implies clear sequential dependencies among the interactions or elements inside a session or a sequence. We show the difference between the session data and the sequence data from a certain user in Table \ref{tab:seq}.    
\begin{small}
\begin{table*}[bp]
 \vspace{-1em}
  \centering
  \caption{\label{tab:seq} A comparison between session data and sequence data}
   \vspace{-0.5em}
   {\footnotesize
    \begin{tabular}{|l|l|c|c|c|p{5cm}|}  
    \hline
     \multicolumn{2}{|c|}{\textbf{Data type}} & \textbf{Boundary} & \textbf{Order} & \textbf{Time interval} & \makebox[5cm][c]{\textbf{Main relations embedded}} \\ \hline
     \multirow{2}{1cm}{Session data} & Unordered session & Multiple & No & Non-identical &Co-occurrence-based dependencies \\\cline{2-6}
                   & Ordered session & Multiple & Yes &  Non-identical &Co-occurrence-based dependencies and sequential dependencies    \\ \hline
     \multicolumn{2}{|c|}{Sequence data} & Single & Yes &  Not included &Sequential dependencies   \\ \hline    
      \end{tabular} 
     }
  \label{tab:data}
  \vspace{-1em}
\end{table*} 
\end{small}

An SBRS aims to predict either the unknown part (e.g., an item or a batch of items) of a session given the known part, or the future session (e.g., the next-basket) given the historical sessions via learning the intra- or inter-session dependencies. Such dependencies usually largely rely on the co-occurrence of interactions inside a session and they may be sequential or non-sequential \cite{hu1937diversifying}. In principle, an SBRS does not necessarily rely on the order information inside sessions, but for ordered sessions, the naturally existing sequential dependencies can be utilized for recommendations. In comparison, an SRS predicts the successive elements given a sequence of historical ones by learning the sequential dependencies among them. Several survey papers focus particularly on SRSs, including sequence-aware recommender systems \cite{quadrana2018sequence}, deep learning for sequential recommendations \cite{fang2019deep} and sequential recommender systems \cite{wang2019sequential}. In this survey, we particularly focus on the area of SBRSs with an emphasis on the unique characteristics of session data together with the corresponding challenges they have brought to SBRSs, and the representative and the state-of-the-art approaches for SBRSs. 


\subsection{A Framework for Organizing SBRS Work} \label{classifi}
The variety of existing work on SBRSs can be generally categorized into three sub-areas fitting a unified categorization framework to reduce the aforementioned inconsistencies and confusion. According to the difference on the recommendation tasks, the sub-areas include \textit{next interaction recommendation}, \textit{next partial-session recommendation}, and \textit{next session recommendation}. Given the known part (i.e., happened interactions) of a session, next interaction recommendation aims to recommend the next possible interaction in the current session by mainly modeling intra-session dependencies. It is usually simplified to predict the next item to interact, e.g., a product to click or purchase. Given the known part of a session, next-partial session recommendation aims to recommend all the remaining interactions to complete the current session, e.g., to predict all the subsequent items to complete a basket given the purchased items in it, by mainly modelling intra-session dependencies. Although less studied, this sub-area is even more practical in real-world cases since a user often does not interact with only one item in the next, but multiple items till 
she finish the whole session. Given the historical sessions, next session recommendation aims to recommend the next session, e.g., next basket, by mainly modeling inter-session dependencies. Sometimes, inter-session dependencies are also incorporated into the first two sub-areas to improve recommendation performance.
A comparison of these sub-areas is presented in Table \ref{tab:task}.    
\begin{small}
\begin{table*}[htp]
 \vspace{-0.8em}
  \centering
  \caption{\label{tab:task} A comparison of different sub-areas in SBRSs}
  \vspace{-0.5em}
   {\footnotesize
    \begin{tabular}{|p{2.3cm}|p{2.4cm}|p{1.8cm}|p{6cm}|}  
    \hline
     \makebox[2.3cm][c]{\textbf{Sub-area}} & \makebox[2.4cm][c]{\textbf{Input}} & \makebox[1.8cm][c]{\textbf{Output}}  & \makebox[6cm][c]{\textbf{Typical research topic}}\\ \hline
    Next interaction recommendation & Mainly known part of the current session & Next interaction (item) & Next item recommendation, next song/movie recommendation, next POI recommendation, next web page recommendation, next news recommendation, etc.\\ \hline
    Next partial-session recommendation & Mainly known part of the current session & Subsequent part of the session & Next items recommendation, session/basket completion\\ \hline
    Next session recommendation & Historical sessions & Next session & Next basket recommendation, next bundle recommendation, etc.\\ \hline
      \end{tabular} 
     }
  \label{tab:data}
  \vspace{-1em}
\end{table*} 
\end{small}
\subsection{Related Surveys}\label{position}
Although many studies have been done in the area of SBRSs, to the best of our knowledge, there are limited comprehensive and systematic reviews to shape this vibrant area and position the existing works as well as the current progress. Although some works have attempted to comprehensively evaluate and compare the performance of existing SBRS algorithms, we have not found any studies which systematically formalize this research field, or comprehensively analyze the unique characteristics of session data and the critical challenges faced by SBRSs. Let alone to provide an in-depth summary of current eﬀorts or detail the open research issues present in the field.  

Several surveys focus on the conventional RSs or the emerging deep learning based RSs. Shi et al. \cite{shi2014collaborative} comprehensively analyzed the recently proposed algorithms for collaborative filtering based RSs and discussed the future challenges in the area. Lops et al. \cite{lops2011content} provided an overview of content-based RSs by summarizing the corresponding representative and the state-of-the-art algorithms and discussing the future trends. Burke et al. \cite{burke2002hybrid} surveyed the landscape of hybrid RSs. Zhang et al. \cite{zhang2019deep} provided a comprehensive review of recent research eﬀorts on deep learning based RSs. In addition, there are also several surveys on SRSs. For instance, Quadrana et al. \cite{quadrana2018sequence} conducted a comprehensive survey on sequence-aware RSs from various aspects, including the recommendation task, the algorithms and evaluations; Fang et al. \cite{fang2019deep} provided a comprehensive survey on deep learning based sequential recommendations from the aspects of algorithms, influential factors, and evaluations; and Wang et al. \cite{wang2019sequential} conducted a brief review on the challenges and progress of SRSs. 

However, there is a lack of an extensive review on SBRSs. Although SBRSs have been partially discussed in \cite{quadrana2018sequence}, this work mainly focused on sequence-aware RSs and only discussed a small proportion of works on SBRSs built on ordered session data, while ignoring SBRSs based on unordered sessions. More importantly, a variety of recent progress has not been covered. To the best of our knowledge, there are only three formally published review papers (include a short one) particularly focusing on SBRSs. To be specific, Ludewig et al. \cite{ludewig2018evaluation}
provided a systematic performance comparison of a number of SBRS algorithms, including Recurrent Neural Network (RNN) based approaches, factorized Markov chain based approaches, and nearest neighbour based approaches. Later, Ludewig et al. \cite{ludewig2019performance} compared the performance of four
neural network based approaches and five conventional approaches based on rule learning or nearest neighbour, which was further extended into a comprehensive empirical study of SBRS algorithms where 12 algorithmic approaches were compared in terms of their recommendation performance \cite{ludewig2019empirical}. 
While focused on the experimental perspective, these studies only covered a few approaches but did not provide a comprehensive review and analysis from the theoretical perspective.  

Given the rising popularity and potential of SBRSs and the steady flow of novel research contributions in this area, a comprehensive survey will be of high scientific and practical
value. This survey seeks to provide a comprehensive review of the current research on SBRSs to bridge these gaps with the aim of supporting further development of SBRSs. As the first attempt, this paper explores the field of SBRSs with an emphasis on the problem statement, analysis of challenges, review of progress and discussion of future prospects.

\section{SBRS Problem Statement }\label{Statement}
An RS can be seen as a system \cite{cao2016non,cao2015coupling}, which consists of multiple basic entities including \textit{users}, \textit{items} and their \textit{behaviours}, e.g., user-item interactions. These basic entities and behaviours form the core constituents of a \textit{session}, which is the core entity in an SBRS. 
\begin{footnotesize}
\begin{table*}[ht]
  \centering
  \caption{\label{tab:natation} Main notations in SBRSs}
   \vspace{-0.6em}
    \begin{tabular}{|p{1.2cm}|p{5.5cm}||p{1.2cm}|p{4.5cm}|}  
    \hline
     \makebox[1.2cm][c]{\textbf{Notation}} & \makebox[5.5cm][c]{\textbf{Description}} & \makebox[1.2cm][c]{\textbf{Notation}}  & \makebox[4.5cm][c]{\textbf{Description}}\\ \hline
    $U$ & A set of users, i.e., $U=\{u_1,u_2,...,u_{|U|}\}$  & $S$ & A set of sessions\\ \hline
    $u_k$ & The $k$-th user of $U$, $1\leq k \leq |U|$ & $s_{j'}$  & The $j'$-th session of $S$, which is a list of interactions\\ \hline
    V & A set of items, i.e., $V=\{v_1,v_2,...,v_{|V|}\}$ & $c$ & A session context   \\ \hline
    $v_i$ & The $i$-th item of $V$, $1 \leq i \leq |V|$ &  $l$ & A list of interactions \\ \hline
    $A$ & A set of actions, i.e., $A=\{a_1,a_2,...,a_{|A|}\}$ & $\bm{v}_i$ & The representation\footnotemark of $v_i$\\ \hline
     $a_{k'}$ & The $k'$-th action of $A$, corresponding to the $k'$-th type of action, e.g., purchase or click &  $\bm{o}_{i^{'}}$ &The representation of $o_{i^{'}}$\\ \hline
    $O$ & A set of interactions, i.e., $O=\{o_1,o_2,...,o_{|O|}\}$ & $\bm{e}_c$ & The representation of $c$ \\ \hline
     $o_{i^{'}}$ & The $i'$-th interaction of $O$, which is a tuple of a user, an item and an action & $\bm{h}_t$ & The hidden state\footnotemark at time step $t$ \\ \hline
    
      \end{tabular} 
  \label{tab:data}
 \begin{tablenotes}
        \footnotesize 
         \item[1] $^{1}$ A representation is often specified as a latent vector; 
         $^{2}$ A hidden state is a latent vector from an RNN (cf. Sec. \ref{RNN aproach}).
 \end{tablenotes}
 \vspace{-1em}
\end{table*} 
\end{footnotesize}
Therefore, we first introduce the definitions and properties of these entities and behaviours, and then define the SBRS problem based on them. These definitions and properties will be further used for the characterization and categorization of SBRSs, etc. The main notations are 
listed in Table \ref{tab:natation}.

\subsection{User and User Properties}\label{user}
A \textit{user} in an SBRS is the subject who takes actions, e.g., clicks, purchases, on items, e.g., products, and receives the recommendation results. 
Let $u$ denote a user and each user is associated with a unique ID and a set of attributes to describe her, 
e.g., the gender of a user, and it has multiple values, e.g., male and female. The attributes of a user could affect the actions she takes on items and further affect the corresponding sessions. For instance, a boy may watch more action movies, leading to more action movies in his watching sessions, while a girl may like to watch more love-story movies. In addition to the explicit attributes that can be obviously observed, some implicit attributes that reflect the user's internal states, e.g., her moods and intentions, may also have a significant impact on her actions. All the users together form the user set, namely $U=\{u_1,u_2,...,u_{|U|}\}$. Note that the user information of a session may not be always available for two reasons: (1) it is not recorded due to the privacy protection; and (2) some users do not log in when interacting with online platforms like 
amazon.com. Consequently, the session becomes anonymous.

\subsection{Item and Item Properties\label{object}}
An \textit{item} in an SBRS is an entity to be recommended, such as a product, e.g., a book, or a service, e.g., a course. Let $v$ denote an item, which is associated with a unique ID and a set of attributes to provide the description information of the item, such as the category and the price of an item. All the items in a dataset form the item set, namely $V=\{v_1,v_2,...,v_{|V|}\}$. 

Usually, items are different in different domains. For instance, in the news recommendation domain, an item is a news article posted on a news website, e.g., a report on artificial intelligence on abc.com; in the e-commerce domain, an item is a product for sale, e.g., an earphone on amazon.com, and in the service industry, an item is a specific service, e.g., the course "Machine Learning" provided by Coursera ( https://www.coursera.org/).   

\subsection{Action and Action Properties}
An \textit{action} is often taken by a user on an item in a session, e.g., clicking an item. Let $a$ denote an action, which is associated with a unique ID and a set of attributes to provide its property information 
e.g., the type of the action, and has multiple values, e.g., click, view, and purchase. Note that some actions may not be associated with specific items, e.g., a search action or a catalog navigation action. But they may still provide useful information to an SBRS as discussed in~\cite{quadrana2018sequence}.  

\subsection{Interaction and Interaction Properties}
Interaction is the most basic unit in sessions. Let $o$ denote an \textit{interaction}, which is a ternary tuple consisting of a user $u$, an item $v$ and the action $a$ taken by $u$ on $v$, namely $o= \langle u, v, a \rangle$. In the case where the user information is not available, the interaction become anonymous, i.e., $o =\langle v, a \rangle$. Moreover, in the case, where there is only one type of actions, e.g., clicks, the interaction $o$ can be further simplified as $o =\langle v \rangle$, namely it only consists of an item. All the interactions together form the interaction set $O$.

\subsection{Session and Session Properties \label{session}}
Session is an important entity in an SBRS. Let $s$ denote a session, which is a non-empty bounded list of interactions generated in a period of continuous time which may be connected with some user- (e.g., user ID) or session-specific (e.g., a session-ID or a cookie) information, i.e., $s=\{o_1, o_2,...,o_{|s|}\}$. Note that here we use the concept "list" instead of "set" to indicate that there may be duplicated interactions in one session. For example, a user listens to a song for multiple times in a listening session. Each session is associated with a set of attributes, 
e.g., the duration of $s$, which have multiple corresponding values, e.g., 20 minutes or 40 minutes. Some other important attributes of a session include the time and the day when the session happens. Next, we discuss five important properties of sessions that may have a great impact on SBRSs. 

\textit{Property 1: session length.} The length of a session is defined as the total number of interactions contained in it. This is a basic property of sessions, which is taken as one of the statistical indicators of experiment data in most literature \cite{hu1937diversifying,wang2017perceiving}. Sessions of different lengths may bring different challenges for SBRSs and thus lead to different recommendation performance. The session characteristics related to session length together with the corresponding challenges for building SBRSs are discussed in detail in Section \ref{Session Length}. 

\textit{Property 2: internal order.}
The internal order of a session refers to the order over interactions within it. Usually, there are different kinds of order flexibility inside different sessions, i.e., no order, flexible order and order. The existence of internal order leads to the sequential dependencies within sessions which can be used for recommendations. The session characteristics related to internal order and the corresponding challenges for building SBRSs are discussed in detail in Section \ref{Internal Order}.        

\begin{figure*}[!t]
\vspace{-0.8em}
	\centering
	\includegraphics[width=.85\textwidth]
	{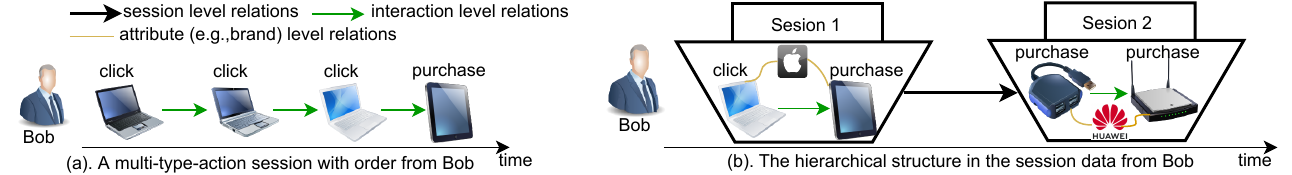}
	\vspace{-1em}
	\caption{Toy examples for a typical session and session data with hierarchical structure}
	\label{fig_session}
	\vspace{-1.5em}
\end{figure*}

\textit{Property 3: action type.}
In the real world, some sessions contain only one type of actions, e.g., purchase, while other sessions may contain multiple types of actions, e.g., click, purchase (cf. Fig. \ref{fig_session} (a)). The dependencies over different types of actions are often different. For instance, the items that are clicked together in a session may be similar or competitive while the items purchased together in one session may be complementary. Therefore, the number of action types in a session determines whether the intra-session dependencies are homogeneous (based on a single type of actions) or heterogeneous (based on multi-type actions), which is important for accurate recommendations. The session characteristics related to action type as well as the corresponding challenges for building SBRSs are discussed in detail in Section \ref{Action Type}.

\textit{Property 4: user information.}
User information of a session mainly refers to the IDs of the users in the session, and sometimes user attributes are also included. In this paper, the property of user information refers to the availability of user information in a session. In the real word, the user information of sessions is given in some cases, while it is not available in other cases (cf. Section \ref{user}) \cite{vasile2016meta,hidasi2015session,wu2019session}. User information plays an important role to connect sessions from the same user happening at different time and thus its availability determines the possibility to model the long-term personalized preference across multiple sessions for a specific user. In practice, SBRSs were initially proposed to handle those anonymous sessions where user information is not available \cite{hidasi2017recurrent}. The session characteristics together with the corresponding challenges for building SBRSs 
are discussed in detail in Section \ref{User Information}.        

\textit{Property 5: session-data structure.} Session-data structure refers to the session-related hierarchical structure consisting of multiple levels \cite{quadrana2017personalizing,Wang2020Jointly}, which intrinsically exists in some session data. For example, the attribute level consists of the attributes of entities (e.g., users, items) in an interaction, the interaction level consists of interactions in each session, and the session level consists of multiple historical sessions from the current user (cf. Fig. \ref{fig_session} (b)). The interaction level is necessary for a session, while the other levels depend on the specific session data. This is because either the attribute information or the historical session information may not be available in all session data. Usually, the number of levels included in a session data set determines the information volume that can be used for recommendations. The session characteristics related to session-data structure as well as the corresponding challenges for building SBRSs are discussed in detail in Section \ref{Session Structure}.

\subsection{The SBRS Problem}\label{SBRS}
From the system perspective, we formalize SBRS by first illustrating its input, output and work mechanism, and then presenting the problem formalization.

\textit{Input.}
The basic input of an SBRS is the partially known session information that is used for recommendations \cite{chen2019dynamic}. According to the specific scenarios, the basic input has three cases: (1) the known part of the current session (i.e., a list of happened interactions), which is the input of the SBRSs modelling intra-session dependencies only for next interaction (item), or next partial-session recommendation (c.f. Section \ref{classifi}); (2) the list of known historical sessions, which is the input of the SBRSs that mainly model inter-session dependencies for next session (e.g., a basket) recommendation; and (3) the combination of the first two, which is the input of the SBRSs modelling both intra- and inter-session dependencies for the recommendation of next interaction, or next partial-session. 

Most of the existing SBRSs mainly take the IDs of users (if not anonymous), items and actions as the input while ignoring their attribute information \cite{hidasi2015session,twardowski2016modelling,liu2018stamp}. Only a minority of studies~\cite{wang2017perceiving,hidasi2016parallel,tavakol2014factored,jannach2017session} assume the attribute information is available and take it as part of the input. In a specific case, the input part of the current session or historical sessions may be anonymous or non-anonymous, ordered or unordered, with single- or multi-type actions. From our observation, most of the existing SBRSs assume the input sessions are ordered and with single-type actions. In an SBRS, the input is usually formalized as a \textit{session context} (also called a context in this paper) conditioned on which the recommendation is performed.

\textit{Output.}
The goal of an SBRS is to make recommendations according to a given session context, i.e., the known session information. Accordingly, the output of an SBRS is a predicted interaction or a predicted list of interactions that happen subsequently in the current session, or the predicted next session that happens following the given historical sessions (cf. Section \ref{classifi}). In an SBRS, the user information of a session is either not available and thus is not predictable (in anonymous sessions) or given by default (in non-anonymous sessions), so it is usually unnecessary to predict the user information. Consequently, each interaction in the output only contains an item and the corresponding action taken on it \cite{tanjim2020attentive}, e.g., an item to purchase or click. Moreover, in most cases where SBRSs are built on single-type-action sessions, the only action type is given by default, e.g., purchase or click, and thus each interaction in the output is further simplified to an item \cite{hu1937diversifying,jing2017neural}. According to the 
specific sub-areas, there are three cases for the output (cf. Section \ref{classifi}): (1) in next interaction recommendation, the output is a list of alternative interactions (items) \cite{tanjim2020attentive,jing2017neural}, ranked by best match as the next interaction (item) in the session; (2) in next partial-session recommendation, the output is a list of interactions (items) to complete the current session; and (3) in next session recommendation, the output is a list of complementary interactions (items) to form the next session \cite{wang2020intentionnet}, e.g., to purchase a basket of complementary products (e.g., milk, bread) to achieve a unified goal (e.g., breakfast). In the last two cases, according to whether the session is ordered or not, the interactions in the list may be ordered or unordered accordingly.

\textit{Work mechanism.}
In principle, the work mechanism of an SBRS is to first learn the comprehensive dependencies among interactions within or/and between sessions and then to utilize the learned dependencies to guide the prediction of the subsequent interactions or sessions to accomplish the recommendation task. Next, we illustrate it in detail in the following problem formalization. 

\textit{Problem formalization.}
There are many different types of SBRSs with their own specific characteristics and formalization. Here we give an abstract-level formalization that is suitable for different SBRSs. Let $l=\{o_1,...,o_j,...,o_n\}$ be a list of $n$ interactions, each of which is composed of an item and the corresponding action taken on it. Recall that in SBRSs built on single-type-action sessions, each interaction is simplified to an item and thus the interaction list $l$ is simplified to an item list $l_v$, i.e., $l_v=\{v_1,...,v_j,...,v_n\} (v_j \in V)$. $L$ is the set of lists, which contains all the possible interaction lists derived from the candidate item set $V$ and action set $A$. $c$ is the input, i.e., a session context, consisting of all the session information used for the recommendation. All the session contexts together form the session context set $C$. Similar to sequence-aware RS~\cite{quadrana2018sequence}, let $f$ be a utility function to return the utility score of a candidate interaction list $l$ for a given session context $c$. An SBRS is to select the recommended interaction list $\hat{l} \in L$ by maximizing the utility score conditioned on the given session context $c$, i.e., 
{\small
\begin {equation}
\centering
\hat{l}= arg\  max\ f(c,l), c \in C,l \in L,
\end {equation} }where the utility function can be specified to multiple forms, e.g., likelihood, conditional probability. The utility function is employed on the interaction list to optimize the candidate list as a whole rather than optimising a single candidate interaction (item). This makes the formalization not only cover all the aforementioned three cases for the output of SBRSs, but it also makes it possible to characterize and evaluate the list as a whole from multiple aspects, e.g., the novelty or diversity of the interactions (items) in the list \cite{quadrana2018sequence}.

\section{Characteristics and Challenges \label{challenge}}

SBRSs are built on session data, and different types of session data are usually associated with different characteristics, which essentially bring different challenges to build SBRSs. Similar to understanding data characteristics and challenges in other data-driven research \cite{dst_cao18}, a deep understanding of the intrinsic characteristics of session data, and the challenges in modelling session data for building SBRSs is fundamental for designing an appropriate SBRS. Therefore, in this section, we systematically illustrate and summarize a variety of characteristics of session data as well as the corresponding challenges caused by each of them in building SBRSs.  

According to each of the session properties introduced in Section \ref{session}, sessions can be divided into different types. For instance, according to the property "session length", sessions can be divided into long sessions, medium sessions and short sessions. Next, we first present different types of sessions categorized by each of their properties, and then discuss the characteristics and challenges associated with each type of sessions. 

\subsection{Characteristics and Challenges Related to Session Length}\label{Session Length}
According to session length, sessions can be roughly categorized into three types: long sessions, medium sessions and short sessions, while the specific definitions for long, medium and short sessions may vary upon the specific data sets. 

\textit{Long sessions.}
A long session contains relatively more interactions, e.g., more than 10. In general, with more interactions, long sessions can provide more contextual information for more accurate recommendations. However, due to the uncertainty of user behaviours, a long session is more likely to contain random interactions \cite{hu1937diversifying} which are irrelevant to other interactions in it. This brings noisy information and thus reduces the performance of recommendations \cite{wang2018attention,wang2019sequential}. Therefore, the first challenge for SBRSs built on long sessions is \textit{how to effectively reduce the noisy information from the irrelevant interactions}. In addition, there are usually more complex dependencies embedded in a long session, e.g., \textit{long-range dependencies} \cite{yuan2019simple} between two interactions that are far from each other in a session or \textit{high-order dependencies} \cite{wang2019sequential} across multiple interactions in a session. Consequently, another challenge for SBRSs built on long sessions is \textit{how to effectively learn complex dependencies for better recommendation performance.}

\textit{Medium sessions.}
Medium sessions usually contain a medium number of interactions, e.g., 4 to 9. From our observations on session data generated from transaction records in e-commerce industry, medium sessions are the most common case \cite{wang2019modeling}. Compared with long and short sessions, a medium session is less likely to contain too many irrelevant interactions while it usually contains the necessary contextual information for Session-Based Recommendation (SBR). Although relatively less challenging, building SBRSs on medium sessions still faces a general challenge, i.e., \textit{how to effectively extract the relevant and accurate contextual information for accurate recommendations.}   

\textit{Short sessions.}
A short session consists of quite limited interactions, e.g., usually less than 4, leading to limited information available for recommendation. For example, in an offline anonymous session consisting of two interactions, the only contextual information that can be utilized to recommend the second interaction (item) is the first interaction in the session. An extreme case is to recommend the first interaction of a session. 
Consequently, the challenge for SBRSs built on short sessions is \textit{how to effectively make recommendations with quite limited contextual information.}

\subsection{Characteristics and Challenges Related to Internal Order}\label{Internal Order}
According to whether there is an order over interactions inside a session or not, sessions can be roughly divided into unordered sessions, ordered sessions and flexible-ordered sessions.

\textit{Unordered sessions.}
An unordered session contains interactions without any chronological order between them, namely, whether an interaction happens earlier or later in the session makes no difference \cite{hu1937diversifying}. For example, the shopping sessions are sometimes unordered since users may pick up a basket of items (e.g., \{bread, milk, eggs\}) without following an explicit order \cite{wang2017perceiving}. In unordered sessions, the dependencies among the interactions are based on their co-occurrence rather than the sequences of them, and thus the generally utilized sequence models are not applicable. Compared with sequential dependencies, co-occurrence based dependencies are usually relatively weak and fuzzy, which are more difficult to learn. 
Furthermore, most of co-occurrence based dependencies among interactions are collective dependencies \cite{tang2018personalized,yuan2019simple}, i.e., several contextual interactions in a session collaboratively lead to the occurrence of the next interaction, which are even harder to capture. Consequently, the challenge for SBRSs built on unordered sessions is \textit{how to effectively learn the relatively weak and fuzzy dependencies among interactions, especially those collective dependencies.}

\textit{Ordered sessions.}
An ordered session contains multiple interactions with strict order, and usually strong sequential dependencies exist among them. For example, the session composed of a sequence of online courses taken by a user is often ordered, since some prerequisite courses must be taken first to gain prior-knowledge for the subsequent ones. Although it is relatively easy to learn the strong sequential dependencies within ordered sessions, it is challenging \textit{to effectively learn the cascaded long-term sequential dependencies which decay gradually with time in long ordered sessions.}

\textit{Flexibly-ordered sessions.}
A flexibly-ordered session is neither totally unordered nor totally ordered, i.e., some parts of the session are ordered while others are not~\cite{tang2018personalized}. For example, a tourist generates a session of check-ins at airport,
hotel, shopping center, bar, and attraction successively. In the session, the airport,
hotel and attraction are actually sequentially dependent, while the shopping center and bar are randomly inserted without any order. Therefore, the complex dependencies inside flexibly-ordered sessions must be carefully considered and precisely learned for accurate recommendation. Consequently, the challenge for SBRSs built on flexibly-ordered sessions comes from \textit{how to effectively learn the complex and mixed dependencies, i.e., sequential dependencies among ordered interactions and non-sequential dependencies among unordered ones.}

\subsection{Characteristics and Challenges Related to Action Type}\label{Action Type}

According to the number of action types included in a session, sessions can be divided into single-type-action sessions and multi-type-action sessions.

\textit{Single-type-action sessions.}
A single-type-action session includes one type of actions only, e.g., clicks of items, and thus only one type of dependencies comes from the same type of actions, which is relatively easy to learn.

\textit{Multi-type-action sessions.}
A multi-type-action session includes more than one types of actions \cite{li2018learning}, leading to multiple types of interactions. For example, in a real-world online shopping session, a user usually first clicks several items for comparison and then purchases one or more items. Thus, there are complex dependencies inside a multi-type-action session \cite{meng2020incorporating}. Specifically, dependencies not only exist over the interactions from the same type (e.g., clicks of items), but also exist over interactions from different types (e.g., clicks and purchases). As a result, a big challenge for SBRSs built on multi-type-action sessions is \textit{how to effectively and accurately learn both the intra- and inter-action type dependencies for accurate recommendations.}

\subsection{Characteristics and Challenges Related to User Information}\label{User Information}
According to whether the user information is available or not, sessions can be divided into non-anonymous sessions and anonymous sessions.

\textit{Non-anonymous sessions.}
A non-anonymous sessions contains non-anonymous interactions with the associated user information, which enables the connections of different sessions generated by the same user at different time. This makes it possible to learn the user's long-term preference as well as its evolution across sessions. However, due to the relative long time-span and preference dynamics, it is quite challenging \textit {to precisely learn the personalized long-term preference over multiple non-anonymous sessions.}

\textit{Anonymous sessions.}
In anonymous sessions, due to the lack of user information to connect multiple sessions generated by the same user, it is nearly impossible to collect the prior historical sessions for the current session. As a result, only the contextual information from the current session can be used for recommendations. Therefore, it is challenging to \textit{precisely capture the user's personalized preference with limited contextual information to provide accurate recommendation.}   

\subsection{Characteristics and Challenges Related to Session-data Structure}\label{Session Structure}
According to the number of levels of structures, session data can be roughly divided into single-level session data and multi-level session data (cf. \ref{session}). Specifically, the interaction level naturally exists in any session data, and thus single-level session data particularly refers to the data including the interaction level only. Multi-level session data refers to the data including the attribute level or/and the session level in addition to the interaction level.   


\textit{Single-level session data.}
A single-level session data set is usually a set of anonymous sessions where each consists of several interactions without attribute information or historical session information. In such a case, only single-level dependencies, i.e., the inter-interaction dependencies within sessions, can be utilized for recommendations. Hence, due to the lack of auxiliary information from other levels, SBRSs built on single-level session data may easily suffer from the cold-start or data sparsity issue \cite{meng2020incorporating}. This leads to the challenge of \textit{how to overcome the cold-start and sparsity issues for accurate recommendations when only the inter-interaction dependencies are available.}    

\begin{tiny}
\begin{table*}[!htb]
  \centering
  \caption{\label{tab:challenge} \begin{footnotesize} Comparison of representative works regarding targeted session type, basic model and application domain 
  \end{footnotesize}}
   \begin{threeparttable}
    \begin{tabular}{|p{2.9cm}|p{1.2cm}|p{0.5cm}|p{0.6cm}||p{3.4cm}|p{1.2cm}|p{0.8cm}|p{0.6cm}|}  
    \hline
      \makebox[2.9cm][c]{\textbf{Work}} &  \makebox[1.2cm][c]{\textbf{ Session type}} &  \makebox[0.5cm][c]{\textbf{Model}}  &  \makebox[0.6cm][c]{\textbf{Domain}} &  \makebox[3.4cm][c]{\textbf{Work}} &  \makebox[1.2cm][c]{\textbf{Session type}} &  \makebox[0.8cm][c]{\textbf{Model}}  &  \makebox[0.6cm][c]{\textbf{Domain}}\\ \hline
     An SBRS based on association rules \cite{mobasher2001effective} &  FO, ST, A, SL &  ARD &  Web page & Attention-gated recurrent network (IARN) \cite{pei2017interacting}  &L, O, ST, NA, SL &RNN, ATT & Video, movie\\ \hline
     
    Access Pattern Approach (APA) \cite{shao2009music} &  UO, ST, NA, SL  & FPM & Music & KNN-GRU4Rec \cite{jannach2017recurrent} & O, ST, A, SL & KNN, RNN &Item\\ \hline
    
    Personalized sequential pattern \cite{yap2012effective} & L\tnote{1}, O, ST, NA, SL & SPM  & Item & Temporal deep semantic structured model (TDSSM) \cite{song2016multi} & O, ST, NA, ML  & MLP, RNN & News    \\ \hline
    
     Item/session KNN \cite{jannach2017recurrent,ludewig2018evaluation} & O/UO, ST, A, SL & NN & Item & List-wise deep neural network \cite{wu2017session} & UO, MT, A, SL & MLP & Item  \\ \hline
    
    Sequence and
    Time Aware Neighbourhood (STAN) \cite{garg2019sequence} & O, ST, A, SL & NN & Item  & DeepPredict \cite{jannach2017session} & UO, MT, NA, ML  & MLP & Fashion   \\\hline
    
    Temporal-Item-Frequency-based User (TIFU)-KNN \cite{hu2020modeling}& UO, ST, NA, SL  & NN &Item & ConvolutionAl Sequence Embedding Recommendation Model (CASER) \cite{tang2018personalized} & FO, ST, NA, SL  & CNN &  Movie, POI  \\ \hline
    
  Page rank and Markov model \cite{eirinaki2005web} & O, ST, A, SL  & MC &Web page & 3D Convolutional Neural Network (3D CNN) \cite{tuan20173d} & O, ST, A, ML & CNN & Item  \\ \hline
  
 Factorized Personalized Markov Chain (FPMC) \cite{rendle2010factorizing} & O, ST, NA, ML & MC, MF & Item & Hierarchical Temporal Convolutional Networks (HierTCN) \cite{you2019hierarchical} & L, O, MT, NA, ML  & TCN & Item\\ \hline
 
 Dynamic emission and transition model \cite{le2016modeling} & O, ST, NA, SL   & HMM & Music, tweet & Session-based Recommendation with
Graph Neural Network (SR-GNN) \cite{wu2019session} & O, ST, A, SL  &GNN & Item  \\ \hline
    
Personalized Ranking Metric
Embedding (PRME) \cite{feng2015personalized} & O, ST, NA, SL & MC, ME & POI & Graph Contextualized Self-Attention Network
(GC-SAN) \cite{xu2019graph} & O, ST, A, SL  &GNN, ATT & Item  \\ \hline

Personalized Markov Embedding (PME) \cite{wu2013personalized} & O, ST, NA, SL   & MC, ME & Music & Target Attentive Graph Neural Network (TAGNN) \cite{yu2020tagnn} & O, ST, A, SL & GNN, ATT & Item  \\ \hline

Latent Markov
Embedding (LME) \cite{chen2012playlist} & O, ST, NA, SL & MC, ME & Music & Multi relational GNN for Session-based Prediction (MGNN-SPred) \cite{wang2020beyond} & O, MT, A, SL   &GNN  & Item  \\ \hline

  FPMC-Localized Regions (LR) \cite{cheng2013you}
 & O, ST, NA, SL   & MF & POI & Full GNN based on Broadly
Connected Session graph (FGNN-BCS) \cite{qiuexploiting} &O, ST, A, ML &GNN & Item \\ \hline

Co-factorization (CoFactor)   \cite{liang2016factorization} & UO, ST, NA, SL  & MF & Item & Full GNN based on Weighted Graph
ATtention layer (FGNN-WGAT) \cite{qiu2019rethinking} & O, ST, A, SL &GNN &Item \\ \hline

Category aware POI recommendation model \cite{liu2013personalized} & O, ST, NA, SL   & MF & POI & Hierarchical Attentive Transaction Embedding (HATE) \cite{Wang2020Jointly} &UO, ST, NA, ML &ATT, DR & Item   \\ \hline

Session-based Wide-In-Wide-Out (SWIWO) networks \cite{hu1937diversifying} & UO, ST, NA, SL   & DR & Item & Dynamic Co-attention Network for SBR (DCN-SR) \cite{chen2019dynamic} & L, O \& UO, MT, NA, ML  & ATT, RNN &Item  \\ \hline
    
 Network-based Transaction Embedding Model (NTEM) \cite{wang2017perceiving} & UO, ST, A, ML   & DR & Item & Encoder-Decoder with attention (EDRec)  \cite{loyola2017modeling} & O, MT, NA, SL &ED, RNN, ATT & Item  \\ \hline   
 
 Meta-Prod2Vec \cite{vasile2016meta} & O, ST, A, ML   & DR & Music & Neural Attentive Recommendation Machine (NARM) \cite{li2017neural} &L, O, ST, A, SL  & ED, RNN ATT  & Item \\ \hline 
 
 Music Embedding Model (MEM) \cite{wang2016learning} & O, ST, NA, ML   & DR & Music & Short-Term Attention/Memory Priority (STAMP) model \cite{liu2018stamp}  &UO, ST, A, SL & ATT, MLP &Item  \\ \hline 
 
 Attention based Transaction Embedding Model (ATEM) \cite{wang2018attention} & UO, ST, A, SL  & DR, ATT & Item & Streaming Session based Recommendation Machine (SSRM) \cite{guo2019streaming}  &O, ST, NA, ML &ATT, MF, RNN & Music, POI   \\ \hline
     
 Hierarchical Representation Model (HRM) \cite{wang2015learning} & UO, ST, NA, ML  & DR & Item & Sequential Hierarchical Attention Network (SHAN) \cite{ying2018sequential} & FO, ST, NA, ML &ATT, DR & Item \\ \hline 
 
 GRU4Rec \cite{hidasi2015session} & O, ST, A, SL  & RNN & Item & Memory-Augmented
  Neural Network (MANN) \cite{chen2018sequential} & FO, ST, NA, ML &MN &Item \\ \hline
 
   GRU4Rec-BPR \cite{hidasi2017recurrent}
   & O, ST, A, SL  & RNN & Item & Memory Augmented Neural model (MAN) \cite{mi2020memory} &FO, ST, A, SL &MN &Item \\ \hline
     
   Improved RNN \cite{tan2016improved}
   & O, ST, A, SL  & RNN & Item & Hierarchical Memory Networks (HMN) \cite{song2019session} &O, ST, A, SL &MN, CNN &Item   \\ \hline

   Hierarchical RNN (HRNN) \cite{quadrana2017personalizing}
   & L, O, ST, NA, ML  & RNN & Item & Collaborative Session-based Recommendation
   Machine (CSRM)  \cite{wang2019collaborative} & O, ST, A, ML & MN, ED &Item \\ \hline  
   
   User-based RNN \cite{donkers2017sequential}
   & O, ST, NA, SL  & RNN, ATT & Movie, music & Multi-temporal-range Mixture Model (M3) \cite{tang2019towards} & L, FO, ST, A, SL  &MM & Movie  \\ \hline  
   
   Dynamic REcurrent bAsket Model (DREAM) \cite{yu2016dynamic}
   & UO, ST, NA, ML  & RNN & Item & Mixture-Channel Purpose Routing Networks (MCPRN) \cite{wang2019modeling} & FO, ST, A, SL &MM, RNN & Item   \\ \hline  
     
  Recurrent Latent Variable network for SBR (ReLaVaR) \cite{chatzis2017recurrent}
   & O, ST, A, SL  & RNN, VI & Item & Intention2basket \cite{wang2020Intention} & FO, ST, A, ML &MM, RNN& Item  \\ \hline

   RNN-latent cross \cite{beutel2018latent}
   & O, ST, NA, ML  & RNN & Video &  Variational Recurrent Model
  (VRM) \cite{wang2018variational} &O, ST, A, SL &GM, RNN& Item  \\ \hline
   
   Parallel RNN (P-RNN) \cite{hidasi2016parallel}
   & O, ST, A, ML  & RNN & Video, item & VAriational SEssion-based Recommendation (VASER) \cite{zhou2019variational}&O, ST, A, SL &GM, RNN& Item  \\ \hline

   Neural survival recommender \cite{jing2017neural} & O, ST, NA, SL  & RNN &Music, movie &  Convolutional Generative Network (NextItNet) \cite{yuan2019simple} &O, ST, A, SL &GM, CNN&Item, music \\ \hline

  Recurrent Recommender Networks (RRN)  \cite{wu2017recurrent} & O, ST, NA, SL & RNN & Video & LIst-wise Recommendation based on Deep RL (LIRD) \cite{zhao2017deep}&O, MT, NA, SL & RL&Item \\ \hline

    Session-aware recommendations with
    neural network \cite{twardowski2016modelling} & L, O, MT, A, SL & RNN & Item & 
    DeepPage \cite{zhao2018deep}& O, MT, NA, SL & RL, RNN& Item \\ \hline

   \multicolumn{8}{|l|}{\textbf{Session type} S: short, M: medium, L: long; O: ordered, UO: unordered, FO: flexible ordered; ST: single-type-action, MT: multi-type-action; A: anonymous, NA:}
   \\
   \multicolumn{8}{|l|}{
   non-anonymous; SL: single-level, ML: multi-level.}\\
   
   \multicolumn{8}{|l|}{\textbf{Model} ARD: association rule discovery, FPM: frequent pattern mining, SPM: sequential pattern mining, NN: nearest neighbour,
   MC: Markov chain, MF: matrix-}\\ 
   
   \multicolumn{8}{|l|}{
    factorization, VI: variational inference, HMM: hidden markov model, ME: metric embedding, DR: distributed representation, RNN: recurrent neural network,} \\
   
    \multicolumn{8}{|l|}{ CNN: convolution neural network, TCN: temporal convolution network, MLP: multi-layer perceptron, GNN: graph neural network, ATT: attention, MN: mem-} \\
    
    \multicolumn{8}{|l|}{ ory network, MM: mixture model, GM: generative model, RL: reinforcement learning, ED: encoder-decoder.}    \\\hline  
   
      \end{tabular} 
       \begin{tablenotes}
      
      \scriptsize
        \item[1] Most of the works do not target sessions of a specific length, only a few ones are experimented on long sessions and thus marked `L'.
    \end{tablenotes}
    
 \end{threeparttable}  
      
  \label{tab:data}
\end{table*} 
\end{tiny}
\textit{Multi-level session data.} Multi-level session data involves a hierarchical structure of at least two levels, i.e., the interaction level plus attribute level and/or session level. In this case, both the dependencies within each level and across different levels would affect the subsequent recommendations. For example, the categories (the attribute level) of several items may have impact on whether these items would be bought together (the interaction level) in one session. Consequently, \textit{how to comprehensively learn the intra- and inter-level dependencies for effective and accurate recommendations} becomes a key challenge for SBRSs built on multi-level session data. 

\subsection{A Summary of Characteristics and Challenges}
In this subsection, we provide a summary of session characteristics and the corresponding challenges. To be specific, a comparison of carefully selected representative and state-of-the-art works on SBRSs regarding their targeted session types, basic model and the application domain is presented in Table \ref{tab:challenge}. Recall that each type of sessions have their own characteristics and challenges, therefore, the session type in Table \ref{tab:challenge} actually reflects the corresponding session characteristics and challenges for SBRSs that are targeted by each work. As a result, Table \ref{tab:challenge} provides a comprehensive overall view on existing works from multiple perspectives. For instance, the second row on the left side in Table \ref{tab:challenge} means that the association rule based SBRS in \cite{mobasher2001effective} mainly targets the Flexible-Ordered (FO), Single-Type-action (ST), Anonymous (A), and Single-Level (SL) sessions with associate rule mining approach for web page recommendations.

\section{Classification and Comparison of SBRS Approaches}\label{Classification}
To provide an overall view of the achieved progress in addressing the challenges introduced in Section \ref{challenge}, we first classify the approaches for SBRSs from the technical perspective, i.e., the involved method or model, in Section \ref{techniqueCategory}, and then compare different classes of approaches in Section \ref{comparison}.

\subsection{A Classification of SBRS Approaches}\label{techniqueCategory}
\begin{figure*}[!t]
	\centering
	\includegraphics[width=.97\textwidth]
	{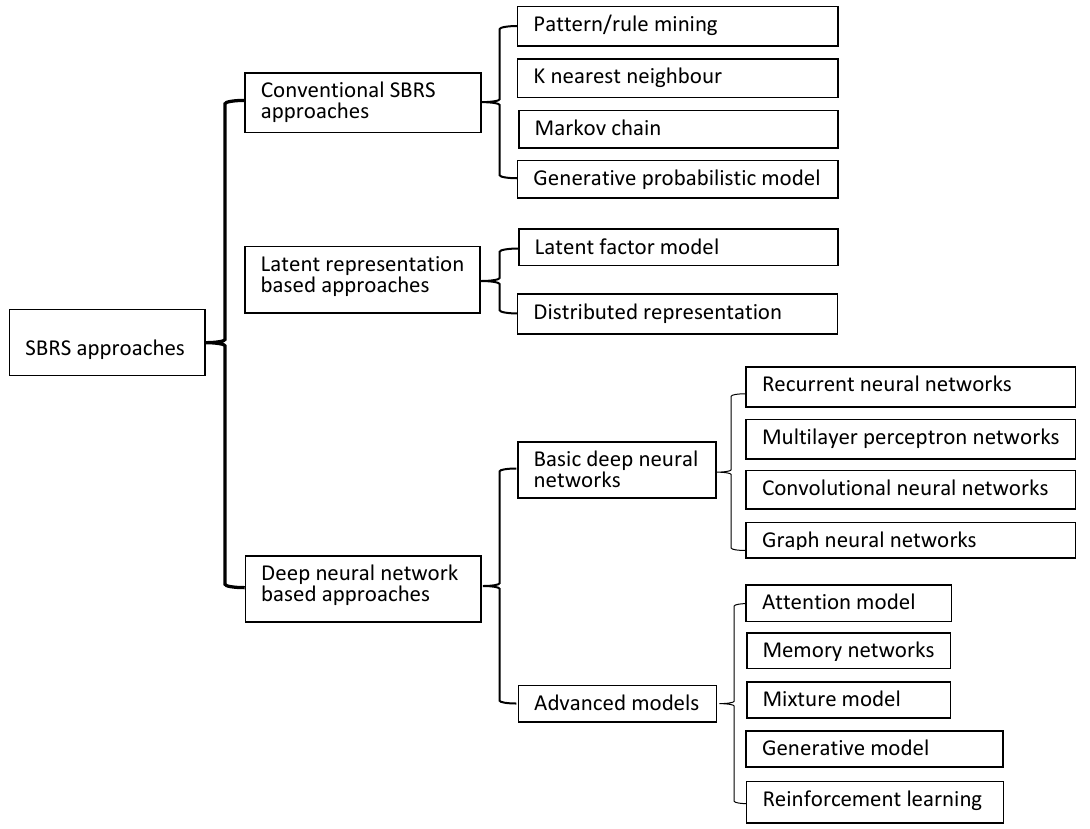}
	\vspace{-1em}
	\caption{The categorization of SBRS approaches}
	\label{fig_class}
	\vspace{-1em}
\end{figure*}

The taxonomy of approaches for SBRS is presented in Fig.  \ref{fig_class}. According to the employed technique, three super-classes of approaches for SBRSs are identified from the literature, i.e., \textit{conventional SBRS approaches, latent representation approaches}, and \textit{deep neural network approaches}. These three super-classes can be further divided into seven classes in total while each super-class contains multiple classes. To be specific, conventional SBRS approaches contain four classes: \textit{pattern/rule-based approaches, KNN-based approaches}, \textit{Markov chain based approaches}, and \textit{generative probabilistic model based approaches}; latent representation approaches contain two classes: \textit{latent factor model } and \textit{distributed representation}; and deep neural network approaches contain two classes: \textit{basic deep neural networks} and \textit{advanced models}. In addition, the class of basic deep neural networks  contains four sub-classes while each corresponds to one basic deep neural network architecture, namely, \textit{recurrent neural networks, multi-perceptron layer (MLP) networks, convolution neural networks} and \textit{graph neural neural networks}. Similarly, the class of advanced model contains five sub-classes while each corresponds to one type of models that are usually utilized for building SBRSs, i.e., \textit{attention models, memory networks, mixture models, generative models} and \textit{reinforcement learning}. Consequently, the existing approaches for SBRSs are classified into three super-classes, eight classes, plus nine sub-classes. As a result, 15 atomic classes of SBRS approaches are obtained, i.e., six classes from conventional SBRS approaches and latent representation approaches, and nine sub-classes from deep neural network approaches. In addition to approaches based on a single technique/model, there are some hybrid approaches which combine multiple techniques/models, e.g., an approach for next-basket recommendations combines Markov chain model and latent factor model~\cite{rendle2010factorizing}. Next, first, a comparison of different classes of approaches will be presented in Section \ref{comparison}, and then each super-class of approaches will be reviewed in Sections \ref{Conventional approaches}, \ref{Latent approaches} and \ref{Deep approaches} respectively.   

\subsection{A Comparison of Different Classes of Approaches }\label{comparison}
Generally speaking, conventional SBRS approaches are relatively simple, straightforward, and easy to understand and implement. Although simple, they are effective in some cases, especially on those simple datasets where the dependencies within or between sessions are obvious and easy to model and capture. Particularly, in a study by Ludewig et al. \cite{ludewig2019performance}, KNN-based approaches, e.g., session-KNN, have achieved superior recommendation accuracies even compared with some deep neural network based approaches, e.g., GRU4Rec, in much less running time on some e-commerce datasets including RETAIL, DIGI\footnote{https://www.dropbox.com/sh/n281js5mgsvao6s/AADQbYxSFVPCun5DfwtsSxeda?dl=0}. In contrast, deep neural network based approaches are usually relatively complex, involve complicated and multi-layer network architecture and 
often require extensive computing. They are generally believed to be more powerful to comprehensively model and capture the complex dependencies, e.g., long-term or high-order dependencies, embedded in complex datasets, e.g., imbalanced or sparse datasets, for more accurate SBR \cite{zhang2019deep}. The superiority of deep neural network based approaches has been verified by a variety of works in recent years, e.g., \cite{hidasi2015session, yu2016dynamic, you2019hierarchical}. In general, latent representation based approaches are a bit more complicated than conventional approaches but less complicated than deep neural network based approaches. Unlike deep neural network based approaches, they usually do not involve a deep network architectures, leading to relatively low computation cost. However, benefiting from their efficient and effective representation learning, they sometimes perform very well. In some studies \cite{liu2018stamp,wang2018attention}, latent representation based approaches can outperform not only some conventional approaches, e.g., Markov chain based approaches \cite{rendle2010factorizing}, but also some deep neural network based ones, e.g., RNN-based approaches \cite{hidasi2015session}.    

As introduced in Section \ref{SBRS}, the work mechanism of SBRSs is to learn the comprehensive dependencies to guide the subsequent recommendations. Therefore, learning dependencies in session data is the key computation task in an SBRS. In addition, recall that most of the challenges in SBRSs can be abstracted to learn the various types of dependencies, e.g., high-order dependencies, embedded in different types of session data, e.g., long sessions, as illustrated in Section \ref{challenge}. Therefore, in order to have a better understanding of how each class of approaches can benefit completing the key computation task and addressing the main challenges in the field of SBRSs, Table \ref{tab:dependency} compares all the 15 atomic classes of approaches regarding the type of dependencies they can learn. For example, the fourth row in Table \ref{tab:dependency} means that Markov chain based approaches mainly capture sequential, short-term, first-order and pointwise dependencies in session data for recommendations. First-order dependency and pointwise dependency refer to the dependency between any two adjacent interactions and that between any pair of interactions, respectively.  

\begin{figure*}[!t]
	\centering
	\includegraphics[width=.7\textwidth]
	{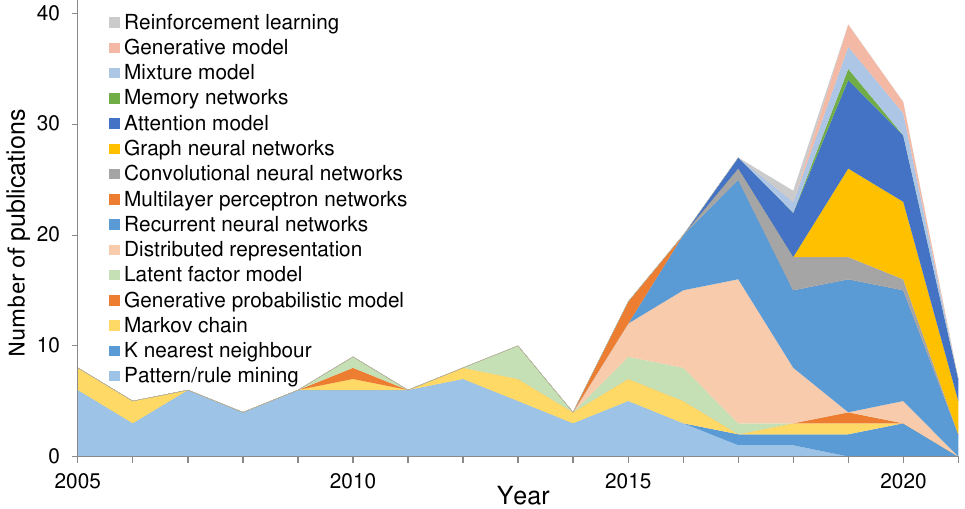}
	\vspace{-0.5em}
	\caption{Number of publications on each class of SBRS per year}
	\label{fig_publication}
\end{figure*}

In addition, a statistic on the number of publications in each atomic class is presented in Fig. \ref{fig_publication}. This result was achieved by manually retrieving and counting the publications on SBRS using Google scholar on 20 March, 2021. We first used the typical keywords "session, recommendation", "next item/basket/POI/song/news/video recommendation" for searching and then manually counted those relevant publications only.  
\begin{footnotesize}
\begin{table*}[htb]
   \vspace{-1em}
  \centering
  \caption{\label{tab:dependency} A comparison of learned dependencies by different classes of approaches}
  \begin{threeparttable}
    \begin{tabular}{|p{3.8cm}|p{2.6cm}|p{1.3cm}|p{2cm}|p{2cm}|}  
    \hline
    \makebox[3.8cm][c]{\textbf{Approach}} & \textbf{Sequential or non-
    sequential} & \textbf{Short- or long-term} & \textbf{First or high-order} & \textbf{Pointwise or collective}  \\ \hline
     
      Pattern/rule mining   & Both\tnote{1}   & Both  & Both & Both 
       \\ \hline
     
      K nearest neighbour & Mainly non-sequential  & Both & Mainly first-order & Both\tnote{2}  \\  \hline
     
     Markov chain  & Sequential &  Short-term  & First-order & Pointwise   \\ \hline
     
     Generative probabilistic model  & Sequential &  Long-term  & Higher-order & Collective   \\ \hline
     
     Latent factor model & Sequential & Short-term & First-order & Pointwise \\ \hline
     
     Distributed representation & Mainly non-sequential & Both & Mainly first-order & Collective \\ \hline
     
     Recurrent neural networks & Sequential & Long-term & High-order & Pointwise \\ \hline

     Multilayer perceptron networks & Non-sequential & Both & First-order & Collective  \\ \hline

     Convolutional neural networks & Mainly sequential & Both & Mainly first-order & Collective   \\ \hline
     
     Graph neural networks & Both & Both & High-order & Pointwise   \\ \hline

     Attention models & Mainly non-sequential & Both & First-order & Mainly pointwise  \\ \hline

     Memory networks & Non-sequential & Both & First-order & Pointwise  \\ \hline

     Mixture models &Both & Both & Both & Both  \\ \hline

     Generative models\tnote{3} &Either & Either & Either & Either\\ \hline
     
     Reinforcement learning & Sequential & Both & High-order & Pointwise \\ \hline
     
    \end{tabular} 
    
    \begin{tablenotes}
      
      \scriptsize
        \item[1] Non-sequential and sequential dependencies are learned by frequent pattern mining and sequential pattern mining respectively.
        \item[2] Item-KNN and session-KNN mainly models pointwise and collective  dependencies respectively. 
        \item[3] The learned dependencies mainly depend on the employed encoder for encoding the input of the generation model.   
    \end{tablenotes}
    
 \end{threeparttable}  
 \vspace{-1.5em}
\end{table*} 
\end{footnotesize}

\section{Conventional SBRS approaches}\label{Conventional approaches}
Conventional approaches for SBRSs utilize the conventional data mining or machine learning techniques, to capture the dependencies embedded in session data for session-based recommendations. Next, we introduce each of the four classes of conventional approaches respectively.

\subsection{Pattern/Rule Mining based SBRSs}\label{pattern}
Generally speaking, there are two types of pattern/rule mining based approaches for SBRS:  (1) \textit{frequent pattern/association rule mining based approaches}, which mine the association rules over different interactions within \textit{unordered sessions} to guide the subsequent recommendations; and (2)  \textit{sequential pattern mining based approaches}, which mine the sequential patterns over sequences of sessions or interactions within \textit{ordered sessions} to guide the subsequent recommendations. This class of approaches can handle single-type-action sessions only in which all the actions are the same in a dataset, so each interaction in a given session is simplified into an item.

\subsubsection{Frequent Pattern/Association Rule Mining based Approaches}\label{Rule-based Approaches}
Frequent pattern/association rule mining based SBRSs mainly contain three steps: (1) frequent pattern or association rule mining, (2) session matching, and (3) recommendation generation. To be specific, given an item set $V$ and the corresponding session set $S$ over $V$, first, a set of frequent patterns $FP=\{p_1,p_2,...,p_{|FP|}\}$ are mined by using pattern mining algorithms like 
FP-Tree \cite{han2000mining}. Then, given a partial session $\hat{s}$ (e.g., a list of chosen items in one session), if an item $\hat{v} \in V  \setminus \hat{s}$ exists so that $ \hat{s} \cup \{ \hat{v} \} \in FP $, then $\hat{v}$ is a candidate item for recommendations. Finally, if the conditional probability $P(\hat{v}|\hat{s})$ is greater than a predefined confidence threshold, then $\hat{v}$ is added into the recommendation list \cite{mobasher2001effective,wang2017inferring}. 

Besides the aforementioned basic framework, there are many variants. For instance, to consider the different significance of different web pages and thus to recommend more useful ones, several methods \cite{forsati2009web, yan2006incorporating} utilized the page-view duration to weight the significance of each page and then incorporated such weight into association rule mining to build \textit{weighted association rule based SBRSs}. With regard to the application 
domain, except for the traditional shopping basket based product recommendations, frequent pattern/association rule based SBRSs are also commonly applied in web page recommendations \cite{moreno2004using}, music recommendations \cite{shao2009music}, and so on.

\subsubsection{Sequential Pattern Mining based Approaches}\label{SequentialRS}
Following a representative work \cite{yap2012effective} falling into this sub-class, here we introduce a typical sequential pattern mining based SBRS built on session level (cf. Section \ref{session}) for next session recommendations. For sequential pattern based mining SBRSs built on the interaction level for next interaction recommendations, please refer to \cite{niranjan2010developing}. Similar to frequent pattern/association rule mining based approaches, sequential pattern mining based SBRSs also contain three steps: (1) sequential pattern mining, (2) sequence matching, and (3) recommendation generation. Specifically, given a sequence set $Q=\{q_1,q_2,...,q_{|Q|}\}$ where $q=\{s_1, s_2,..., s_{|q|}\}$ is a sequence of sessions from the same user ordered according to the timestamp, first, a set of sequential patterns $SP=\{p_1,p_2,...,p_{|SP|} \}$ is mined on $Q$. Then, given a user $u$'s sequence $q_u=\{s_1, s_2,...,s_g \}$, for any sequential pattern $p \in SP$, if the last session $s_g$ of $q_u$ belongs to $p$, i.e.,  $p=\{s_1, s_2, ...,s_g,s_r... \}$, then $p$ is a relevant pattern for this specific recommendation and the items after $s_g$ in $p$, like items in $s_r$, are candidate items. For each candidate item $\hat{v}$, its support is the sum of the support of all relevant patterns:
{\small
\begin {equation}
\centering
 supp(\hat{v})=\sum_{s_g \in q_u, s_g\in p, \hat{v}\in s_r, s_r\in p, p\in SP}   supp(p).
\end {equation} }Finally, those candidate items with the top support values are recommended to user $u$.

Except for the basic framework described above, there are various extensions for sequential pattern mining based SBRSs. A typical example is to utilize the user-related weighted sequential pattern mining for personalized recommendations, where each sequence is assigned a weight based on its similarity to those past sequences of the target user \cite{song2014personalized}.
Another extension is to build a hybrid RS by combining sequential pattern mining and collaborative filtering to consider both the users' dynamic individual patterns and their general preference \cite{choi2012hybrid,liu2009hybrid}. Regarding the application domain, shopping basket based product recommendations \cite{yap2012effective} and web page recommendations \cite{niranjan2010developing} are two typical applications of sequential pattern based SBRSs.

\subsection{K Nearest Neighbour based SBRSs}
K Nearest Neighbour (KNN) based approaches for SBRS are proven to be simple but effective \cite{ludewig2018evaluation}. In principle, a KNN-based SBRS first finds out the $K$ interactions or sessions that are most similar to the current interaction or session respectively from the session data. Then, it calculates a score for each candidate interaction based on the similarity to indicate its relevance to the current interaction as the guidance of recommendations. For the same reason as mentioned in the first paragraph in Section \ref{pattern}, each interaction is simplified as an item in this class of approaches. According to whether the similarity is actually calculated between items or sessions, KNN-based approaches for SBRSs can be divided into \textit{item-KNN} and \textit{session-KNN}. 

\subsubsection{Item-KNN}
Given the current session context, an item-KNN based SBRS recommends those $K$ items most similar to the current item in terms of their co-occurrence in other sessions as the next choice. Technically, each item is encoded into a binary vector where each element indicates whether the item occurs (set to "1") in a specific session or not (set to "0"). Consequently, the similarity between items can be calculated on their vectors with a certain similarity measure, like cosine similarity \cite{ludewig2018evaluation}.  

\subsubsection{Session-KNN\label{sessionKNN}}Given the current session context $c$, a session-KNN based SBRS first calculates the similarity between $c$ and all other sessions to find the set $N(c)$ of its $K$ neighbour sessions, and then calculates the score of each candidate item $\hat{v}$ w.r.t. $c$ based on the similarity:  
{\small
\begin {equation}
\centering
   score(\hat{v})= \sum_{s_{nb} \in N(c)} sim(c, s_{nb})\cdot 1_{s_{nb}}(\hat{v}),
\end {equation} }where $sim$ is a kind of similarity measures and $1_{s_{nb}}(\hat{v})$ is an indicator function which returns $1$ if $\hat{v}$ occurs in $s_{nb}$ and $0$ otherwise. 

Compared with item-KNN, session-KNN considers the whole session context rather than just the current item in the session context, and thus can capture more information for more accurate recommendations. Other similar works include an improved session-KNN which takes into account the readily available sequential and temporal
information from sessions \cite{garg2019sequence}, and a hybrid approach that combines session-KNN and GRU4Rec (i.e., an RNN-based SBRS) using a weighted combination scheme \cite{jannach2017recurrent}. In addition, a user-KNN approach built on users' session information was also proposed for next-basket recommendation \cite{hu2020modeling}.

\subsection{Markov Chain based SBRSs}\label{MarkovApproach}
Markov chain based SBRSs adopt Markov chains to model the transitions over interactions within or between sessions to predict the probable next interaction(s) or session given a session context \cite{shani2005mdp}. According to whether the transition probabilities are calculated based on explicit observations or latent space, Markov chain based approaches can be roughly divided into \textit{basic Markov chain based approaches} and \textit{latent Markov embedding based approaches}.

\subsubsection{Basic Markov Chain based Approaches}

A basic Markov chain based SBRS usually contains four steps:  (1) calculating the transition probabilities over a sequence of interactions, (2) predicting the transition paths over interactions, (3) matching the session context to the predicted paths, and (4) making recommendations based on the matching result \cite{eirinaki2005web}. Note that, in most cases, the interactions here are simplified to items.

To be specific, a Markov chain model is defined as a set of tuples $\{ST, \bm{P_t}, P_0\}$, where $ST$ is the state space including all the distinct interactions, $\bm{P_t}$ is the $m*m$ one-step transition probability matrix between $m$ distinct interactions, and $P_0$ is the initial probability of each state in $ST$. First, the first-order transitional probability from interaction $o_i$ to $o_j$ is defined as: 
{\small
\begin {equation}
\centering
{P_t}(i,j)=P(o_i \rightarrow o_j)= \frac{ freq(o_i \rightarrow o_j)}{\sum_{o_t } freq(o_i \rightarrow o_t)}.
\end {equation} }Second, a transition path, e.g., $\{o_1 \rightarrow o_2 \rightarrow o_3\}$, is predicted by estimating its probability by using the first-order Markov chain model:
{\small
\begin {equation}
\centering
P(o_1\rightarrow o_2\rightarrow o_3)=P(o_1)* P(o_2|o_1)*P(o_3|o_2).
\end {equation} }Then, given a session context consisting of a sequence of interactions, the paths with high probabilities are chosen as the reference paths. Finally, if the session context occurs in a reference path, those items occurring after it in this path are put into the recommendation list. 

Except for the basic Markov chain based SBRS introduced above, there are many variants. 
For example, Zhang et al. \cite{zhang2007efficient} combined first- and second-order Markov model together to make more accurate web page recommendations. Le et al. \cite{le2016modeling} developed a hidden Markov model based probabilistic model for next item recommendations. 
Rendle et al. \cite{rendle2010factorizing} factorized the transition probability matrix to estimate those unobserved transitions among interactions. 

\subsubsection{Latent Markov Embedding based Approaches}
Different from the basic Markov chain based SBRSs which calculate the transition probabilities based on the explicit observations directly, Latent Markov Embedding (LME) based SBRSs first embed the Markov chains into an Euclidean space and then calculate the transition probabilities between interactions based on their Euclidean distance \cite{chen2012playlist}. In this way, they can derive the unobserved transitions and thus solve the data sparsity issue in limited observed data. Formally, each interaction $o$ is represented as a vector $\bm{o}$ in a $d$-dimensional Euclidean space, and the transition probability $P(o_{i}\rightarrow o_{j})$ is assumed to be negatively related to the Euclidean distance $|| \bm{o}_{i}- \bm{o}_{j} ||_2$ between  $o_{i}$ and $o_{j}$. Accordingly, the probability of a transition path $pa = \{o_1\rightarrow o_2\rightarrow,...,\rightarrow o_{|pa|}\}$ can be defined based on Markov model:
{\small
\begin {equation}
\centering
P(\{o_1\rightarrow o_2\rightarrow,...,\rightarrow o_{|pa|}\})=\prod_{i=2}^{|pa|} P(o_{i-1}\rightarrow o_i)=\prod_{i=2}^{|pa|} \frac{e^{-|| \bm{o}_{i}- \bm{o}_{i-1} ||_2^2 }} {\sum_{o_t} e^{-|| \bm{o}_{t}- \bm{o}_{{i-1}} ||_2^2 }}.
\end {equation} }
\indent To generate personalized recommendations, Wu et al. \cite{wu2013personalized} proposed a Personalized Markov Embedding (PME) model which maps both users and items into an Euclidean space where the user-item distance and item-item distance reflect the corresponding pairwise relationship. Further, Personalized Ranking Metric Embedding (PRME) was proposed to first project each item into a low dimensional Euclidean latent space, and then use the Metric Embedding algorithm to effectively compute transitions between items in a Markov chain model. Intuitively, the Euclidean distances measure the probabilities of transitions \cite{feng2015personalized}.

\subsection{Generative Probabilistic Model based SBRSs}
Generative probabilistic model based approaches generally first infer the latent taxonomy (e.g., topics or genres) of items (e.g., songs) in sessions and then learn the transitions among these latent taxonomies within or between sessions. Afterwards, they predict the next latent taxonomy using the learned transitions. Finally, they further predict specific items as the next item conditional on the predicted latent taxonomy of items. Usually, the latent topic model is utilized to infer the latent taxonomies and the transitions among them. Representative studies include music recommendation based on latent
topic sequential patterns~\cite{hariri2012context}, and playlist generation with statistical models on music-listening sessions~\cite{zheleva2010statistical}.

\subsection{Comparison of Conventional
SBRS Approaches}
After providing the main idea and key technical details of each class of conventional approaches for SBRSs, we present a comparison and summary of those approaches in this subsection. Specifically, in Table \ref{tab:ComConvention}, we compare these three classes of approaches in terms of their applicable scenarios, i.e., for which type of session data that an approach is suitable, pros, cons and the typical works. An empirical comparison on prediction accuracy of the first three classes of conventional SBRS approaches was conducted on seven datasets from domains including retail, music and news. The result shows KNN-based approaches especially the session-KNN achieve superior performance~\cite{jarv2019predictability}.         
\begin{footnotesize}
\begin{table*}[hb]
   \vspace{-1.5em}
  \centering
  \caption{\label{tab:ComConvention} A comparison of different classes of conventional approaches for SBRSs}
   \vspace{-0.5em}
    \begin{tabular}{|p{1.5cm}|p{2.8cm}|p{2.6cm}|p{3.4cm}|p{1.8cm}|}  
    \hline
      \makebox[1.5cm][c]{\textbf{Approach}}   & \makebox[2.8cm][c]{\textbf{Applicable scenario}}      & \makebox[2.6cm][c]{\textbf{Pros}} &  \makebox[3.4cm][c]{\textbf{Cons}} & 
     \makebox[1.8cm][c]{\textbf{Typical work}}
      \\ \hline
     
      Pattern/rule mining based SBRSs   & Simple, balanced and dense, ordered or unordered sessions  & Intuitive, simple and effective on session data where dependencies are easy to learn  & Information loss, cannot handle complex data (e.g., imbalanced or sparse data) & \cite{forsati2009web},\cite{mobasher2001effective},\cite{moreno2004using},
      \cite{niranjan2010developing},\cite{shao2009music},\cite{yap2012effective}
       \\ \hline
     
      KNN based SBRSs & Simple, ordered or unordered sessions  & Intuitive, simple and effective, quick response & Information loss, hard to select $K$, limited ability for complex sessions (e.g., noisy sessions) & \cite{garg2019sequence},\cite{hu2020modeling},\cite{jannach2017recurrent},
      \cite{ludewig2018evaluation} \\  \hline
     
     Markov chain based SBRSs  & Short and ordered sessions with short-term and low-order dependencies &  Good at modelling short-term and low-order sequential dependencies & Usually ignore long-term and higher-order dependencies, the rigid order assumption is too strong & \cite{chen2012playlist},\cite{eirinaki2005web},\cite{feng2015personalized},
     \cite{le2016modeling},\cite{rendle2010factorizing},\cite{wu2013personalized},
     \cite{zhang2007efficient}  \\ \hline
     
      Generative probabilistic SBRSs  & Ordered sessions with high-order dependencies &  Good at modelling high-order and collective dependencies & Computation cost is relatively high & \cite{hariri2012context}, \cite{zheleva2010statistical}  \\ \hline

    \end{tabular} 
    \vspace{-1.6em}
\end{table*} 
\end{footnotesize}

\section{Latent representation approaches for SBRS\lowercase{s}}\label{Latent approaches}
Latent representation approaches for SBRSs first build a low-dimensional latent representation for each interaction within sessions with shallow models. The learned informative representations encode the dependencies between these interactions, and then will be utilized for the subsequent session-based recommendations. According to the utilized techniques, latent representation approaches can be roughly classified into latent factor model based approaches and distributed representation based approaches.

\subsection{Latent Factor Model based SBRSs}\label{FactorizationApproach}

Latent factor model based SBRSs first adopt factorization models, e.g., matrix factorization, to factorize the observed transition matrix over interactions (items) into their latent representations, and then utilize the resultant latent representations to estimate the unobserved transitions for the subsequent session-based recommendations. To be specific, first, a transition tensor $\mathcal{B}^{|U| \times |O| \times |O| }$ can be built using the observed session data, where each entry $b_{k,i,j}$ indicates the transition probability from interaction $o_i$ to $o_j$ under user $u_k$. Then, a general linear factorization model, e.g., Tucker Decomposition, is used to factorize $\mathcal{B}$:
{\small
\begin {equation}\label{eq C}
\centering
\hat{\mathcal{B}}= \mathcal{C}o \times  \bm{U} \times  \bm{O}_{i} \times  \bm{O}_{j},
\end {equation} }where $\mathcal{C}o$ is a core tensor, $\bm{U}$ is the latent representation matrix for users while $\bm{O}_{j}$ and $\bm{O}_{k}$ are the latent representation matrix for the last interactions and the current interactions respectively. 

To alleviate the negative effect of the sparse transitions observed for $\mathcal{B}$, a special case of Canonical Decomposition \cite{bandelt1992canonical} is used to transfer Eq (\ref{eq C}) into the modelling of pairwise interactions: 
{\small
\begin {equation}
\centering
\hat{b}_{k,i,j}=<\bm{u}_k,\bm{o}_{i}> + <\bm{o}_{i},\bm{o}_{j}> + <\bm{u}_k,\bm{o}_{j}>,
\end {equation} }where $\bm{u}_k$, $\bm{o}_{i}$ and $\bm{o}_{j}$ are the latent representation vector of user $u_k$, the last interaction $o_i$ and the current interaction $o_j$ respectively \cite{rendle2010factorizing}. Here, interactions are usually simplified to items. 

In addition to the above defined latent factor model based SBRS, i.e., Factorized Personalized Markov Chain (FPMC) model, there are many other variants. For instance, Cheng et al. \cite{cheng2013you} extended FPMC into FPMC-LR by adding a constraint to limit user movements into a localized region to make it more consistent with the real-world tourism cases for next POI recommendations. A co-factorization model, CoFactor, was proposed to jointly decompose the user-item interaction matrix and the item-item co-occurrence matrix with shared latent factors for items to capture both the users' individual preference and the item transition patterns \cite{liang2016factorization}. Some other similar works \cite{liu2013personalized,lian2013collaborative} utilize the matrix factorization model to learn the transitions of preferences from one location category to another to provide location recommendations.

\subsection{Distributed Representation based SBRSs}

Distributed representation based SBRSs generally learn the distributed representations of interactions (usually specified to items, sometimes users are also incorporated) with a shallow neural network structure to map each interaction into a low-dimensional latent space. In most cases, the shallow neural network structure is similar to Skip-gram model \cite{pennington2014glove} or CBOW model \cite{mikolov2013exploiting} in the natural language processing domain. As a result, the intra- or inter-session dependencies are encoded into the distributed representations, which are then used for session-based recommendations. 

Specifically, a shallow neural network embeds a user $u_k$ and an item $v_i$ into a latent distributional vector respectively using the logistic function $\delta (\cdot)$ for nonlinear transformation \cite{hu1937diversifying}: 
{\small
\begin {equation}
\centering
\bm{u}_k=\delta(\bm{W}^u_{:,k}),
\end {equation} }
{\small
\begin {equation}
\centering
\bm{v}_i=\delta(\bm{W}^v_{:,i}),
\end {equation} }where $\bm{W}^u \in \mathbb{R}^{d \times |U|}$ and $\bm{W}^v \in \mathbb{R}^{d \times |V|}$ are the user and item embedding matrices respectively, and the $k^{th}$ column of $\bm{W}^u$ corresponds to user $u_k$.

In addition to the basic latent representation based SBRS introduced above, there are a number of variants. Wang et al. \cite{wang2017perceiving} designed a shallow network to embed the ID and features of each item simultaneously to build a compound item representation to tackle the cold-start item issues; some similar works include \cite{vasile2016meta,wang2016learning}. To attentively learn the relevance scales of different interactions in the session context w.r.t. the next choice, attention mechanism is incorporated into the representation learning process \cite{wang2018attention}. In other related works \cite{wan2015next,wang2015learning}, a hierarchical representation of each basket is learned for next-basket recommendations. 

\subsection{Comparison of Latent Representation
based SBRS Approaches}

After providing the main idea and the basic technical details of each class of latent representation approaches for SBRSs, we present a comparison and summary of these approaches. Specifically, in Table \ref{tab:ComLatent}, we compare the two classes of approaches in terms of their applicable scenarios, i.e., for which kind of session data that an approach is suitable, pros, cons and the typical works.     
\begin{footnotesize}
\begin{table*}[htb]
 \vspace{-0.8em}
  \centering
  \caption{\label{tab:ComLatent} A comparison of different classes of latent representation
  approaches for SBRSs}
   \vspace{-0.5em}
    \begin{tabular}{|p{1.5cm}|p{2.8cm}|p{2.2cm}|p{3.3cm}|p{2.3cm}|}  
    \hline
     \makebox[1.5cm][c]{\textbf{Approach}}   & \makebox[2.8cm][c]{\textbf{Applicable scenario}}      & \makebox[2.2cm][c]{\textbf{Pros}} &  \makebox[3.3cm][c]{\textbf{Cons}} &
     \makebox[2.3cm][c]{\textbf{Typical work}}  \\ \hline
     
      Latent factor model   &  Dense, ordered session data & Relatively simple and effective   &  Suffer from data sparsity, cannot capture higher-order and long-term dependencies &\cite{cheng2013you},\cite{lian2013collaborative},\cite{liang2016factorization},
      \cite{liu2013personalized},\cite{rendle2010factorizing},\cite{shani2005mdp}
      
       \\ \hline
     
      Distributed representation & Unordered session data  & Simple and efficient, strong encoding capability & Hard to model ordered or heterogeneous sessions (e.g., noisy sessions) & \cite{greenstein2017session},\cite{hu1937diversifying},\cite{li2017learning},\cite{wan2015next},
      \cite{wang2015learning},\cite{wang2017perceiving},\cite{wang2018attention} \\  \hline
     
    \end{tabular} 
    \vspace{-1em}
\end{table*} 
\end{footnotesize}

\section{Deep neural network approaches for SBRS\lowercase{s}}\label{Deep approaches}
Deep neural network approaches for SBRSs mainly take advantage of the powerful capabilities of deep neural networks in modelling the complex intra- and inter-session dependencies for recommendations. According to the utilized basic framework, deep neural network approaches can be roughly divided into basic deep neural network approaches, each of which involves one type of a basic neural network architecture, e.g., Recurrent Neural Networks (RNN), and advanced models, each of which involves a certain advanced mechanism or model, e.g., attention model.

\subsection{Basic Deep Neural Network based SBRSs}

According to the utilized network architecture, basic deep neural network approaches can be divided into RNN-based approaches, Multi-Layer Perceptron (MLP) based approaches,  Convolutional Neural Networks (CNN) based approaches and Graph Neural Networks (GNN) based approaches. 


\subsubsection{Recurrent Neural Networks (RNN)\label{RNN aproach}}

Benefiting from their intrinsic advantages for modeling sequential dependencies, RNN-based approaches dominate deep neural network approaches for SBRSs. This is because the order assumption has been applied to the interactions in a majority of session datasets in the literature. 
Particularly, an RNN-based SBRS first models each ordered session context as a sequence of interactions within the context. In such a way, it takes the last hidden state of the RNN modeling the context as the context representation. Then, the RNN-based SBRS takes the context representation as the input to predict the next interaction to complete the recommendation task. In those RNN-based SBRSs where the inter-session dependencies are considered, the representation of a sequence of historical sessions is first learned in a similar way and then incorporated for recommendations.  

We introduce a representative RNN-based SBRS called GRU4Rec which is built on Gated Recurrent Units (GRU) \cite{hidasi2015session}, as an example to illustrate the work mechanism of RNN-based SBRSs. To be specific, an RNN is built to model the session context consisting of a sequence of interactions. 
First, the embedding $\bm{o}_t$ of the $t^{th}$ interaction $o_t$ in the context is taken as the input of the $t^{th}$ time step of the RNN. Then, an RNN unit, i.e., GRU, is used to update the hidden state $\bm{h}_{t}$ at the $t^{th}$ time step by absorbing information from both the last hidden state $\bm{h}_{t-1}$ and the current candidate state $\hat{\bm{h}_t}$ by using an update gate $\bm{z}_t$. 
{\small
\begin {equation}
\centering
\bm{h}_t=(\bm{1} - \bm{z}_t)\bm{h}_{t-1} + z_t\hat{\bm{h}_t},
\end {equation} }where  $z_t$ and $\hat{\bm{h}_t}$ are computed by Eqs (\ref{zt}) and (\ref{ht}) given below respectively.
{\small
\begin {equation} \label{zt}
\centering
\bm{z}_t= \sigma(\bm{W}_z\bm{o}_t + \bm{X}_z\bm{h}_{t-1}),
\end {equation} } 
{\small
\begin {equation} \label{ht}
\centering
\hat{\bm{h}_t}= tanh(\bm{W}_h\bm{o}_t + \bm{X}_h(\bm{r}_t \odot \bm{h}_{t-1})),
\end {equation} }where $\odot$ denotes Hadamard product and the reset gate $\bm{r}_t$ is given below:
{\small
\begin {equation}
\centering
\bm{r}_t=\sigma(\bm{W}_r\bm{o}_t + \bm{X}_r\bm{h}_{t-1}), \end {equation} }where $\sigma$ is the activation function which can be specified to be sigmoid function. $\bm{W}$ and $\bm{X}$ are the corresponding weighting matrices. 

In this way, a session context $c$ composed of $|c|$ interactions can be modeled by an RNN with $|c|$ units. Finally, the hidden state $\bm{h}_{|c|}$ from the last time step is used as the representation $\bm{e}_c$ of $c$ for the prediction of the next interaction \cite{hidasi2015session}.

In addition to the basic GRU4Rec, there are also many variants. To improve GRU4Rec, Tan et al. \cite{tan2016improved} adopted data augmentation via sequence preprocessing and embedding dropout to enhance the training process and reduce overfitting respectively. 
Quadrana et al. \cite{quadrana2017personalizing} further improved GRU4Rec by proposing a hierarchical RNN model to capture both intra- and inter-session dependencies for more reliable next item(s) recommendations. Specifically, a two-level GRU-based RNN is designed: the session-level GRU models the sequence of items purchased within each session and generates recommendations for next item(s), while the user-level GRU models the cross-session information transfer and then provides personalized information to the session-level GRU by initializing its hidden state. Another similar work is Inter-Intra RNN (II-RNN) proposed by Ruocco et al. \cite{ruocco2017inter}. In \cite{donkers2017sequential}, the authors designed a unique user-based GRU model which incorporates user characteristic to generalize personalized next item recommendations. Furthermore, there are also RNN-based SBRSs built on basic RNN units, for example, the Dynamic REcurrent bAsket Model (DREAM) \cite{yu2016dynamic} learns a dynamic representation of a user at each time step using an RNN built on basic RNN units for next basket recommendations. 


There are also other variants which incorporate (1) variational inference into RNN to handle the uncertainty in sparse session data and simultaneously enhance the model's scalability on large real-world datasets for recommendations \cite{chatzis2017recurrent,christodoulou2017variational}; (2) side information like item features and contextual factors like time and location into RNN to improve the recommendation performance \cite{hidasi2016parallel,beutel2018latent}; (3) time decay or attention mechanism into RNN to discriminate the intra-session dependencies and thus achieve more precise recommendations \cite{bogina2017incorporating,pei2017interacting}; and (4) traditional models like factorization machines or neighbourhood models to make up the drawbacks of RNN-only models \cite{twardowski2016modelling,jannach2017recurrent}. There are other similar RNN-based SBRSs \cite{jing2017neural,hidasi2017recurrent,wu2017recurrent,gabriel2019contextual}.

\subsubsection{MultiLayer Perceptron (MLP) networks}

MLP-based approaches are usually applied to learn an optimized combination of different representations to form a compound representation of session context for the subsequent recommendations. Different from RNN, MLP is mainly suitable for unordered session data due to the lack of capability to model sequence data. Specifically, in the work of Wu et al. \cite{wu2017session}, an MLP layer is utilized to connect the representations of different parts of a session context to export a unified and compound representation $\bm{e}_c$ for context $c$:  
{\small
\begin {equation}
\centering
\bm{e}_c=\sigma(\bm{W}_c\bm{e}_{c_c} + \bm{W}_v\bm{e}_{c_v}),
\end {equation} }where $\bm{e}_{c_c}$, $\bm{e}_{c_v}$ are the representation of the sub session context containing "click" actions and the sub session context containing "view" actions respectively. $\bm{W}_c$ and $\bm{W}_v$ are the corresponding weight matrices to fully connect each of the representations to the hidden layer of MLP. 

In addition, Jannach et al. \cite{jannach2017session} applied MLP to learn an optimized combination of different factors like "reminders", "item popularity" and "discount" as a compound session-based feature for next-item recommendations. Song et al. \cite{song2016multi} employed an MLP layer to combine both a user's long-term static and short-term temporal preferences for making more accurate next-item recommendations. 

\subsubsection{Convolutional Neural Networks (CNN)}\label{CNN}
CNN are another good choice for SBRSs for two reasons: (1) they relax the rigid order assumption over interactions within sessions, which makes the model more robust; and (2) they have high capabilities in learning local features from a certain area and relationships between different areas in a session to effectively capture the union-level collective dependencies embedded in session data. In principle, a CNN-based SBRS first utilizes the filtering and pooling operations to better learn an informative representation for each session context and then uses the learned representation for the subsequent recommendations \cite{yuan2020future}. To be specific, given a session context $c$ consisting of $|c|$ interactions, an embedding matrix $\bm{E} \in \mathbb{R}^{d \times |c|}$ of $c$ can be constructed by first mapping each interaction in $c$ into a $d$-dimensional latent vector and then puting all the vectors together into a matrix. Afterwards, in a horizontal convolutional layer, the $m^{th}$ convolution value $\alpha_m^x$ is achieved by sliding the $x^{th}$ filter $\bm{F}^x$ from the top to the bottom on $\bm{E}$ to interact with its horizontal dimensions:
{\small
\begin {equation}
\centering
   \alpha_m^x=\phi_\alpha(\bm{E}_{m:m+h-1} \odot \bm{F}^x),
\end {equation} }where $\phi_\alpha$ is the activation function for the convolutional layer.

Then the final output $\bm{e}_c \in \mathbb{R}^z$ from the $z$ filters is obtained by performing the max pooling operation on the convolution result $\bm{\alpha}^x=[\alpha_1^x , \  \alpha_2^x,...,\alpha_{|c|-h+1}^x]$ to capture the most significant features in the session context:
{\small
\begin {equation}
\centering
   \bm{e}_c=max\{max(\bm{\alpha}^1), max(\bm{\alpha}^2), ...,max(\bm{\alpha}^z)\}.
\end {equation}}
\indent Finally, $\bm{e}_c$ is treated as the representation of the session context $c$ and is used for subsequent recommendations \cite{tang2018personalized}.

Some variants include a 3D CNN model \cite{tuan20173d} built for SBRS, which jointly models the sequential patterns in click session data and the item characteristics from item content features, and a CNN model \cite{park2017deep} to accumulate long-term user preferences for generating personalized recommendations. Furthermore, Temporal Convolutional Networks (TCN) were utilized to 
model the interactions within sessions to predict the next interaction \cite{you2019hierarchical}.

\subsubsection{Graph Neural Networks (GNN) \label{GNN}}
In recent years, GNN have shown great expressive power in modeling the complex relations embedded in graph structured data by introducing deep neural networks into graph data~\cite{wang2020modelling,wang2021graph}. To benefit from this power, some researchers have introduced GNN to model the complex transitions within or between sessions to build better-performing SBRSs. First, given a dataset containing multiple sessions, it is transferred to a graph $\mathcal{G}$ by mapping each session into a chain on the graph. Each interaction $o$ in a session serves as a node $n$ in the corresponding chain where an edge $e$ is created to connect each pair of the adjacent interactions in the session. Then, the constructed graph is imported into GNN to learn an informative embedding for each node (interaction) by encoding the complex transitions over the graph into the embeddings. Finally, these learned embeddings are imported into the prediction module for session-based recommendations. According to the specific model architecture of GNN, GNN approaches for SBRSs can be generally divided into three classes: Gated Graph Neural Networks (GGNN), Graph Convolutional Networks (GCN) and Graph ATtention networks (GAT).      

\textit{Gated Graph Neural Networks (GGNN) for SBRSs.}
In GGNN-based SBRSs, first, a directed graph is constructed based on all the historical ordered sessions, where the direction of each edge indicates the order of adjacent interactions within sessions. Then, each session graph, i.e., a chain (sub-graph) for each session, is processed successively by GGNN to obtain the embedding $\bm{n}_i$ of node $n_i$, namely the embedding of the corresponding interaction $o_i$. Finally, after all the session graphs are processed, the embeddings of all the interactions are obtained, which are then used to construct the embedding of the session context for recommendations. Particularly, in GGNN, a Gated Recurrent Unit (GRU) is used to learn the embedding of each node in a session graph by updating the embedding recurrently. Specifically, the embedding (also called hidden state) $\bm{h}^t_i$ of node $n_i$ at step $t$ is updated by the previous hidden state of itself and its neighbourhood nodes, i.e., $\bm{h}^{(t-1)}_i$ and $\bm{h}^{(t-1)}_{j}$, 
{\small
\begin {equation}
\centering
\bm{h}^t_i = GRU (\bm{h}^{(t-1)}_i, \sum_{ {n_j} \in N(n_i)}  \bm{h}^{(t-1)}_{j},  \bm{A}),
\end {equation}}where $N(n_i)$ is the set of neighbourhood nodes of $n_i$ in the session graph, and $\bm{A}$ is the adjacency matrix built on the session graph.
After multiple iterations until a stable equilibrium is reached, the hidden state at the final step of node $n_i$ is taken as its embedding $\bm{n}_i$. Session-based Recommendation
with Graph Neural Networks (SR-GNN) \cite{wu2019session} is the pioneering work which introduced GNN into SBRSs and is claimed to have achieved superior performance, compared with non-GNN approaches including Short-Term Attention/Memory Priority Model (STAMP) \cite{liu2018stamp}, Neural Attentive Recommendation Machine (NARM) \cite{li2017neural} and an RNN-based approach built on GRU, named GRU4Rec \cite{hidasi2015session}. But it may not always outperform conventional methods, as discussed in~\cite{ludewig2019empirical}. Other representative approaches falling into this stream include (1) Graph Contextualized Self-Attention Network
(GC-SAN) \cite{xu2019graph}, which utilizes both GNN and self-attention mechanism to learn local dependencies and long-range dependencies respectively, for session-based recommendations, and (2) Target Attentive GNN (TAGNN) \cite{yu2020tagnn} which first learns item embedding with GNN and then
attentively activates diﬀerent user interests with respect to varied
target items, for session-based recommendations. 

\textit{Graph Convolutional Networks (GCN) for SBRSs.}
Different from GGNN-based SBRSs, GCN-based SBRSs mainly utilize the pooling operation to integrate information from node $n_i$'s neighbourhood node $n_j$ in the graph to help with the update of the hidden state of $n_i$ as shown below:
{\small
\begin {equation} 
\centering
\hat{\bm{h}^t_i}= pooling ( \{  \bm{h}^{(t-1)}_{j}, n_j \in N(n_i)\}),
\end {equation}}where $N(n_i)$ is the set of neighbourhood nodes of node $n_i$. Different specific pooling operations including mean pooling and max pooling can be utilized, depending on the specific scenarios. Afterwards, the integrated neighbourhood information can be incorporated into the iterative update of the hidden state of node $n_i$ \cite{wang2020beyond}
:
{\small
\begin {equation} 
\centering
\bm{h}^t_i=  \bm{h}^{(t-1)}_i + \hat{\bm{h}^t_i}. 
\end {equation} }
\indent Finally, when a stable equilibrium is reached, the last hidden state of node $n_i$ is taken as its embedding $\bm{n}_i$.  

\textit{Graph ATtention networks (GAT) for SBRSs.} 
GAT-based SBRSs mainly utilize attention mechanism to attentively integrate the information from the neighbourhood nodes of node $n_i$ in a session graph to update its hidden state in each attention layer \cite{qiu2019rethinking}:
{\small
\begin {equation} 
\centering
\bm{h}^t_i= attention ( \{  \bm{h}^{(t-1)}_{j}, n_j \in N(n_i)\}),
\end {equation}}where $\bm{h}^t_i$ is the hidden state of node $n_i$ in the $t^{th}$ attention layer. Here $attention$ is a general attention module and can be specified to different operations including self attention, multi-head attention, etc. In principal, the operations in $attention$ can be divided into two steps (cf. Section \ref{atten}): (1) calculating the importance weights of each neighbourhood node, and (2) aggregating the hidden states of neighbourhood nodes according to their importance weights. Finally, once the forward propagation of multiple attention layers
is completed, the hidden state of each node $n_i$ in a session graph at the final layer is taken as its embedding $\bm{n}_i$.

Typical works falling into this class include Full Graph Neural Network (FGNN), which learns the
inherent order of the item transition patterns in sessions with a multiple
Weighted Graph ATtention (WGAT) network \cite{qiu2019rethinking}, another FGNN based on Broadly Connected
Session graph to attentively exploit information both within and between sessions \cite{qiuexploiting}, and Shortcut Graph ATention (SGAT) to effectively propagate information along shortcut connections with attention mechanism~\cite{chen2020handling}.

\subsection{Advanced Model based SBRSs}
In addition to the aforementioned four classes of basic deep neural network approaches for SBRSs, there are also advanced approaches that are built on some advanced models or algorithms, including attention models, memory networks, mixture models, generative models and reinforcement learning. Usually, these advanced models or algorithms are combined with some basic approaches like distributed representation learning or RNN to construct more powerful SBRSs. 

\subsubsection{Attention Models\label{atten}}
Attention-based SBRSs introduce the attention mechanism \cite{vaswani2017attention} to discriminatively exploit different elements, i.e., interactions or/and sessions, in a session context to build an informative session context representation for accurate recommendations. With the incorporation of attention mechanism, an SBRS is able to emphasize those elements that are more relevant to the next interaction or session and reduce the interference of the irrelevant ones in a session context. Generally, an attention model mainly contains two steps: attention weight calculation and aggregation. Next we introduce how an attention model learns a context representation for next interaction recommendations when the context includes the known part of the current session only (cf. Section \ref{classifi}). For 
contexts including historical sessions, their representations can be learned in a similar way.

Step 1: given the embedding $\bm{o}_i$ of interaction $o_i$ in session context $c$ of next interaction $o_{tg}$, attention model calculates the weight $\beta_{tg, i}$ of $o_i$ to indicate its relevance scale w.r.t $o_{tg}$, which is usually performed by a softmax function \cite{wang2018attention}:
{\small
\begin {equation}
  \beta_{tg,i}= \frac{exp(e({\bm{o}_i}))} 
  {\sum_{o_j\in c }  exp(e(\bm{o}_j)) }, 
\end {equation} }where $e({\bm{o}_i})$ is a utility function, which can be specified as the inner product between a learnable weight vector $\bm{w}$ and $\bm{o}_i$. Sometimes, $\bm{o}_{tg}$ is also taken as an input of the utility function to make the learned weight more sensitive to the target interaction $o_{tg}$.

Step 2: the embeddings of all interactions in the session context $c$ are aggregated with the learned weights to construct the embeddeding $\mathbf{e}_{c}$ for $c$: {\small
\begin {equation}
\centering
 \mathbf{e}_{c}= aggregate(\{\bm{o}_i, \beta_{tg, i}, \  o_i \in c\}), 
\end {equation} }where $aggregate$ is an aggregation function which is often specified as a weighted sum. The context embedding is then fed into the prediction module for generating recommendations.   

In addition to the aforementioned basic attention model, a series of variants have been proposed for improving the performance of session-based recommendations. For example, a hierarchical attention model was proposed to attentively integrate both a user's historical sessions and the current session to capture her long- and short-term preferences for accurate session-based recommendations \cite{ying2018sequential,Wang2020Jointly}. Similarly, a co-attention network was designed to better explore the correlations between a user's current interaction and the interactions from historical sessions, for more accurate session-based recommendations \cite{chen2019dynamic}. It should be noted that attention models are usually integrated into other basic approaches, including encoder-decoder \cite{loyola2017modeling}, distributed representation learning \cite{wang2018attention}, RNN \cite{li2017neural} and GNN \cite{qiuexploiting}, to enhance their capabilities for recommendations. In particular, the attention-enhanced GNN, i.e., GAT, has been introduced in Section \ref{GNN}. Other representative approaches for SBRSs that utilize attention models include the Short-Term Attention/Memory Priority (STAMP) model \cite{liu2018stamp}, the self attention model \cite{zhang2019next} and the soft attention model \cite{guo2019streaming}.  

\subsubsection{Memory Networks}
A memory network based SBRS introduces a memory network to capture the dependency between any interaction in the session context and the next interaction directly by introducing an external memory matrix. Such matrix stores and updates the information of each interaction in a secession context more explicitly and dynamically to keep the most relevant and important information for the recommendation task.  

To be specific, a memory network based SBRS mainly consists of two major components: \textit{a memory
matrix} that maintains the embeddings of interactions in a session context $c$, and \textit{a controller} that performs operations
(including reading and writing) on the matrix \cite{chen2018sequential}. Suppose $\mathbf{M}^c$ is the memory matrix to store the embeddedings of the recent interactions in $c$, where each column corresponds to the embedding of one interaction. After an interaction $o_i$ happens in a session and is added into $c$, $\mathbf{M}^c$ will be updated accordingly to maintain the information of the recent interactions by writing the embedding $\mathbf{o}_i$ of $o_i$ into it:
{\small
\begin {equation}
\centering
 \mathbf{M}^c \gets write(\mathbf{M}^c, \mathbf{o}_i), 
\end {equation} }where $write$ stands for the write operation, and it can be specified as one of various writing processes, including the Least Recently Used Access (LRUA) \cite{santoro2016meta}. 

During the prediction, the relevant information is carefully read from the maintained memory matrix to build the embedding $\mathbf{e}_c$ of the session context $c$:
{\small
\begin {equation}
\centering
 \mathbf{e}_c =  read(\mathbf{M}^c, \mathbf{o}_{tg}), 
\end {equation}}where $\mathbf{o}_{tg}$ is the embedding of the next interaction $o_{tg}$ to be predicted, and it is considered during the reading process to read the information more relevant to $o_{tg}$. The read operation can be specified to multiple forms, and a typical one is to use the aforementioned attention mechanism (cf. Section \ref{atten}) to attentively read the information from the memory matrix. 

In addition to the basic memory network based SBRS introduced above, some advanced variants have been proposed for better modeling sessions and making recommendations. For instance, two parallel memory modules, i.e., an Inner Memory Encoder (IME) and an Outer Memory Encoder (OME), were proposed by Wang et al. \cite{wang2019collaborative} to model the current session and neighbourhood sessions respectively to build more informative embedding for a session context. Song et al. \cite{song2019session} proposed hierarchical memory networks to model a user's item-level and feature-level preferences simultaneously for better preforming SBR. Other typical works include the short-term attention/memory priority model for SBR \cite{liu2018stamp}, and the memory augmented neural model for incremental SBR \cite{mi2020memory}.

\subsubsection{Mixture Models}
A mixture model based SBRS mainly builds a compound model containing multiple sub-models to take the advantage of each one to comprehensively model the various complex dependencies embedded in session data. Usually, each sub-model excels at modeling a certain type of dependencies, e.g., low-order or higher-order dependencies. In principle, a mixture model based SBRS performs two main steps: (1) learn different types of dependencies using different sub-models, and (2) carefully integrate the learned dependencies for accurate SBR. 

Representative mixture model based SBRSs include neural Multi-temporal range Mixture Model (M3), which combines different kinds of encoders to capture short- and long-term dependencies in a session respectively for accurate recommendations \cite{tang2019towards}, and Mixture-channel Purpose
Routing Networks (MCPRN), which employs multiple recurrent networks to model the intra-session dependencies under a user's different shopping purposes \cite{wang2019modeling}.

\subsubsection{Generative Models}
Generally speaking, approaches based on generative models for SBRSs make recommendations by generating the next interaction(s) or the next session via a carefully designed generation strategy. In this way, the recommendation procedure better approaches a user's online shopping behaviours in the real word, where items are often picked up step by step to form a shopping basket \cite{wang2020Intention}. To be specific, given a session context $c$ as the prior information, a list of interactions (items) $l$ is generated to serve as the recommendation list:
{\small
\begin {equation}
\centering
 l =  generate(c), 
\end {equation}}where $generate$ stands for a generation process, which can be specified as one of the various forms including probabilistic
generative models \cite{ye2012exploring}.

Representative generative model based SBRSs include \textit{NextItNet} \cite{yuan2019simple} where a probabilistic
generative model was devised to generate a probability distribution over the candidate items; \textit{Intention2Basket} model \cite{wang2020Intention} where a utility-based generator was designed to generate a candidate session with the maximum utility to best fulfill a user's shopping intentions; the \textit{Variational Recurrent Model
(VRM)} where a stochastic generative process of
sessions was specified \cite{wang2018variational}; and \textit{VAriational SEssion-based Recommendation (VASER)} which utilized a non-linear probabilistic method for
Bayesian inference to perform SBR \cite{zhou2019variational}.

\subsubsection{Reinforcement Learning (RL)}
Reinforcement learning approaches for SBRSs generally model the interactions between a user and an RS in a session as a Markov Decision Process (MDP). Note that here the interaction particularly refers to the conversation between a user and an RS. For instance, first, an RS recommends an item to a user who provides some feedback on it, and then the RS recommends the subsequent item according to the user's feedback to better fit her preference. An RL-based SBRS aims to learn the optimal recommendation strategies via recommending trial-and-error items and receive reinforcements for these items from users’ feedback \cite{zhao2017deep}. In this way, an RL-based SBRS is able to continuously update its strategies during the interactions with users until reaching the optimal one that best fits the users' dynamic preferences. Moreover, the expected long-term
cumulative reward from users is considered during the optimization of the strategies. 

Following the work of Zhao et al. \cite{zhao2017deep}, we formalize a basic RL-based SBRS. First, the following five key concepts are defined in an RL-based SBRS. \textit{State space} $Sa$, where a state $sa_t = \{sa_t^1,...,sa_t^{m'} \} \in Sa$ is defined as the previous $m'$ items with which a user
interacted before time $t$. \textit{Action space} $Ac$, where an action $ac_t = \{ac_t^1,...,ac_t^{n'}\} \in Ac$ is to recommend a list of $n'$ items to a user at time $t$ based on the current state $sa_t$. \textit{Reward} $Re$: after the RS takes an action $ac_t$ at the state $sa_t$, it receives
immediate reward $Re_t$ according to the user’s feedback. \textit{Transition probability} $Tp(sa_{t+1}|sa_t,ac_t)$
defines the probability of the state transition from $sa_t$ to $sa_{t+1}$ when the RS takes an action $ac_t$. \textit{Discount factor} $df$: $df\in[0, 1]$ defines the discount factor when we measure the present value of the future reward. Therefore, SBR can be formalized to find
a recommendation policy $\pi: Sa$ → $Ac$ to maximize the
cumulative reward for an RS given the historical MDP, i.e., $(Sa, Ac, Re, Tp, df)$.

There are three main steps in an RL-based SBRS. The first step is to calculate the state-specific weight parameters by mapping state $sa_t$ to a weight matrix $\bm{W}_t$:
{\small
\begin {equation}
\centering
 f_t: sa_t \rightarrow \bm{W}_t.
\end {equation}}The second step is to calculate the score of each candidate item using the score function $f_{s}$ and then select items with the highest score for recommendations:
{\small
\begin {equation}
\centering
 score(v_i) = f_{s}(\bm{v}_i, \bm{W}_t).
\end {equation}}
\begin{footnotesize}
\begin{table*}[htb]
  \centering
   \vspace{-1.2em}
  \caption{\label{tab:ComDeep} A comparison of different classes of deep neural network
  approaches for SBRSs}
   \vspace{-0.8em}
    \begin{tabular}{|p{0.8cm}|p{0.7cm}|p{2.8cm}|p{3.2cm}|p{2.8cm}|p{1.8cm}|}  
    \hline
     \multicolumn{2}{|c|}{\textbf{Approach}}  & \makebox[2.8cm][c]{\textbf{Applicable scenario}}  & \makebox[3.2cm][c]{\textbf{Pros}}  & \makebox[2.8cm][c]{\textbf{Cons}}  & \makebox[1.8cm][c]{\textbf{Typical work}}   \\ \hline

      \multirow{4}{0.8cm}{Basic deep neural networ-ks} & RNN & Long and rigidly ordered sessions & Model long-term and high-order sequential dependencies & The rigid order assumption is too strong for session data & \cite{bogina2017incorporating},\cite{chatzis2017recurrent},\cite{hidasi2015session},
      
      \cite{hidasi2016parallel},\cite{pei2017interacting},\cite{quadrana2017personalizing},
      
      \cite{ruocco2017inter},\cite{tan2016improved},\cite{yu2016dynamic}
      \\\cline{2-6}
      & MLP & Unordered sessions, sessions with multi-aspects (e.g., static and dynamic features) to be combined  & A simple structure, project sparse features to dense ones, learn the combination of different parts &  Cannot model complex sessions, e.g., ordered, heterogeneous sessions & \cite{cheng2016wide},\cite{jannach2017session},\cite{song2016multi},
      \cite{wu2017session} \\\cline{2-6}
      & CNN & Flexible-ordered, heterogeneous or noisy sessions & Robust, no rigid order assumption, capture the union-level collective dependency & Relatively high complexity & \cite{park2017deep},\cite{tang2018personalized},\cite{tuan20173d},
      
      \cite{you2019hierarchical},\cite{yuan2020future}
      \\\cline{2-6}
      & GNN & Complex sessions with complex transitions, e.g., repeat interactions & Model the complex transitions among interactions & Complex and costly & \cite{qiuexploiting},\cite{qiu2019rethinking},\cite{wang2020beyond},
      \cite{wu2019session},\cite{xu2019graph},\cite{yu2020tagnn}
      \\ \hline

      \multirow{5}{0.8cm}{Advanc-ed 
      models} & Atten-tion & Heterogeneous, noisy, or long sessions & Identify and highlight important information & Cannot capture sequential information &
       \cite{guo2019streaming},\cite{li2017neural},\cite{liu2018stamp},
      
      \cite{loyola2017modeling},\cite{Wang2020Jointly},\cite{wang2018attention}, \cite{ying2018sequential},\cite{zhang2019next}
      \\\cline{2-6}
      & Memo-ry & Long, incremental or noisy sessions & Dynamically store the latest information  & Cannot capture sequential information & \cite{chen2018sequential},\cite{mi2020memory},\cite{santoro2016meta},
      \cite{song2019session},\cite{wang2019collaborative} \\\cline{2-6}
      & Mixtu-re & Heterogeneous, noisy sessions & Model different types of dependencies, e.g., long and short term dependencies & Relatively complex and costly& \cite{tang2019towards},\cite{wang2019modeling} \\\cline{2-6}
      & Gener-ative & Dynamic, incremental sessions & Close to the practical session formation & Complex & \cite{wang2020Intention},\cite{wang2018variational},\cite{yuan2019simple} \\\cline{2-6}
      & RL & Dynamic, incremental sessions & Interactive process, consider the future effect of actions  & Hard to simulate the interactive environment & \cite{hu2017playlist},\cite{zhao2018deep},\cite{zhao2017deep} 
       \\  \hline
    \end{tabular} 
    \vspace{-1em}
\end{table*} 
\end{footnotesize}The final step is to calculate the action value $E(sa_t, ac_t)$ of the potential action $ac_t$, i.e., to recommend the selected item, to judge whether $ac_t$ matches the current state $sa_t$ or not \cite{zhao2017deep}. Usually, the following optimal action-value function $E^{*}(sa_t ,ac_t)$, namely the maximum expected return achievable by the optimal policy \cite{zhao2018deep}, is used:
{\small
\begin {equation}
\centering
 E^{*}(sa_t ,ac_t) = \mathbb{E}_{sa_{t+1}} [Re_t + df\  max_{ac_{t+1}} E^{*}(sa_{t+1} ,ac_{t+1})|sa_t,ac_t ].
\end {equation}}Subsequently, the recommendation strategies are optimized by minimizing the error between the action value of the ground truth action and that of trialed actions. 

Typical works on RL-based SBRSs include LIst-wise Recommendation framework based on Deep reinforcement learning (LIRD) \cite{zhao2017deep} which learns recommendation strategies for list-wise recommendations; a similar work called DeepPage for page-wise recommendations \cite{zhao2018deep}; and Reinforcement Learning with
Window for Recommendation (RLWRec) \cite{hu2017playlist} where a state compression method was proposed to capture an enormous state space for play-list recommendations.

\subsection{Comparison of Deep Neural Network based SBRS Approaches}

After introducing the main ideas and the key technical details of deep neural network approaches for SBRSs, we present a comparison and summary of these approaches. In particular, in Table \ref{tab:ComDeep}, we compare the two classes of deep neural network approaches, including nine sub-classes, in terms of their applicable scenarios, the typical example of pros and cons, and the typical works.  

\section{SBRS Applications, Algorithms and Datasets}
\subsection{SBRS Applications}
SBRSs are widely applied in a variety of real-word domains and scenarios to benefit both customers and businesses. A summary of SBRS applications is presented in Table~\ref{tab:application}. Generally speaking, these applications can be grouped into (1) \textit{conventional applications}, e.g., next-item recommendation in E-commerce, and (2) \textit{emerging applications}, e.g., next-treatment recommendation in healthcare.     

According to whether to recommend a product, a content or a service, we can organize SBR into product recommendation, content recommendation and service recommendation. The conventional applications involve all these three classes while the emerging applications mainly involve service recommendation (cf. Table~\ref{tab:application}). From our observation, most of the existing works on SBRS focus on the conventional applications, especially the E-commerce domain, e.g., to recommend the next item~\cite{yap2012effective} or next basket of items~\cite{wang2020Intention} on an online shopping platform (e.g., amazon.com).    
\begin{footnotesize}
\begin{table*}[!htbp]
\vspace{-1.2em}
  \centering
  \caption{\label{tab:application} A summary of SBRS applications}
   \vspace{-1em}
    \begin{tabular}{|p{1.4cm}|p{1.8cm}|p{2.6cm}|p{3.8cm}|p{2.4cm}|}  
    \hline
     \multicolumn{2}{|c|}{\textbf{Category}}  & \makebox[2.6cm][c]{  \textbf{Application domain}}  &\makebox[3.8cm][c]{\textbf{Application scenario}} & \makebox[2.4cm][c]{\textbf{Typical work}}  \\ \hline
     
      \multirow{3}{0.8cm}{Conventional application} & Product recommendation & E-commerce & Next-item/basket recommendation & \cite{hidasi2015session},\cite{hu1937diversifying},\cite{jannach2017recurrent},\cite{ludewig2018evaluation},
      \cite{wang2015learning},\cite{wang2020Intention},\cite{wu2019session},\cite{yap2012effective}
      \\\cline{2-5}
      & Content recommendation & Media, entertainment &Next news/web-page/song/movie /video recommendation&\cite{eirinaki2005web},\cite{jing2017neural},\cite{mobasher2001effective},\cite{pei2017interacting},
      \cite{song2016multi},\cite{tang2019towards},\cite{vasile2016meta},\cite{zhao2017deep}  \\\cline{2-5}
      & Service recommendation & Tourism & Next-POI recommendation & \cite{cheng2013you},\cite{liu2013personalized} 
      \\ \hline

      \multirow{2}{0.8cm}{Emerging application}   &\multirow{2}{0.8cm}{Service recommendation} & Finance & Next-trading recommendation  & \cite{fister2021two},\cite{xiong2018practical}
      \\\cline{3-5}
      &  & Healthcare & Next-treatment recommendation  & \cite{haas2019using}  \\\hline
       
    \end{tabular} 
    \vspace{-1em}
\end{table*} 
\end{footnotesize}In addition, it is not uncommon that SBRSs are applied to other conventional domains, e.g., next news/web-page recommendation in media domain~\cite{song2016multi}, next song/movie/video recommendation in entertainment domain~\cite{zhao2017deep}, and next-POI recommendation in tourism domain~\cite{liu2013personalized}.

Compared with the prosper of conventional applications of SBRS, the emerging applications of SBRS are just in their early stage. However, the applications of SBRS in emerging domains including finance and healthcare are promising and deserve to be further explored. For example, SBRS is of great potential to recommend next trading strategies or portfolios to an investor per her investment goals and context in a financial market and to suggest personalized treatments~\cite{xiong2018practical} on a patient according to her health conditions, past treatments and medical treatment protocols~\cite{haas2019using}.

\subsection{Algorithms and Datasets for SBRSs}
The source code of most of the representative SBRS algorithms is publicly accessible. We summarize the open-source code of algorithms for SBRSs built on different models for various tasks in Table~\ref{tab:link} to facilitate the access for empirical analysis.

\begin{footnotesize}
\begin{table*}[htb]
   \vspace{-1em}
  \centering
  \caption{\label{tab:link} A list of representative open-source SBRS algorithms}
  \vspace{-1em}
  \begin{threeparttable}
    \begin{tabular}{|p{2cm}|p{1.4cm}|p{2cm}|p{1.5cm}|p{5.2cm}|}  
    \hline
    \makebox[2cm][c]{\textbf{Algorithm}} &  \makebox[1.4cm][c]{\textbf{Task}} &  \makebox[2cm][c]{\textbf{Utilized model}} &  \makebox[1.5cm][c]{\textbf{Venue}} &  \makebox[5.2cm][c]{\textbf{Link}}  \\ \hline
     
      TBP~\cite{guidotti2017market}   & Next basket   & Pattern mining  & ICDM 2017 & \scriptsize{\url{https://github.com/GiulioRossetti/tbp-next-basket}}
       \\ \hline
     
       UP-CF~\cite{faggioli2020recency}  & Next basket  & KNN & UMAP 2020 & \scriptsize{\url{https://github.com/MayloIFERR/RACF}} \\  \hline
     
     FPMC~\cite{rendle2010factorizing} & Next basket & Markov chain  & WWW 2010 & \scriptsize{\url{https://github.com/khesui/FPMC}}   \\ \hline
     
     HRM~\cite{wang2015learning} & Next basket &  Distributed representation  & SIGIR 2015 & \scriptsize{\url{https://github.com/chenghu17/Sequential_Recommendation}} \\ \hline
     
     DERAM~\cite{yu2016dynamic} & Next basket & RNN & SIGIR 2016 & \scriptsize{\url{https://github.com/yihong-chen/DREAM}} \\ \hline
     
     Beacon~\cite{le2019correlation} & Next basket & RNN & IJCAI 2019 & \scriptsize{\url{https://github.com/PreferredAI/beacon}} \\ \hline
     
     TIFUKNN~\cite{hu2020TIFUKNN} & Next basket & KNN & SIGIR 2020 & \scriptsize{\url{https://github.com/HaojiHu/TIFUKNN}}\\ \hline

     AR~\cite{ludewig2018evaluation} & Next item & Association rule & UMUAI 2018 & \scriptsize{\url{https://github.com/rn5l/session-rec}} \\ \hline

     BPR-MF~\cite{rendle2012bpr,ludewig2018evaluation} & Next item & Latent factor & UAI 2009 & \scriptsize{\url{https://github.com/rn5l/session-rec}} \\ \hline
     
     IKNN~\cite{jannach2017recurrent} & Next item & KNN & RecSys 2017 & \scriptsize{\url{https://github.com/rn5l/session-rec}} \\ \hline
     
      SKNN~\cite{jannach2017recurrent} & Next item & KNN & RecSys 2017 & \scriptsize{\url{https://github.com/rn5l/session-rec}} \\ \hline

     FOSSIL~\cite{he2016fusing} & Next item & Latent factor & ICDM 2016 & \scriptsize{\url{https://github.com/rn5l/session-rec}}\\ \hline

     SMF~\cite{ludewig2018evaluation} & Next item & Latent factor & UMUAI 2018 & \scriptsize{\url{https://github.com/rn5l/session-rec}}  \\ \hline

     GRU4Rec~\cite{hidasi2015session,hidasi2017recurrent} &Next item & RNN & ICLR 2016 & \scriptsize{\url{https://github.com/rn5l/session-rec}}\\ \hline
     
     STAMP~\cite{liu2018stamp}  &Next item & Attention & KDD 2018 & \scriptsize{\url{https://github.com/rn5l/session-rec}} \\ \hline
     
     NARM~\cite{li2017neural} & Next item & Attention, RNN & CIKM 2017 & \scriptsize{\url{https://github.com/rn5l/session-rec}} \\ \hline
     
   SR-GNN~\cite{wu2019session} & Next item & GNN & AAAI 2019 & \scriptsize{\url{https://github.com/CRIPAC-DIG/SR-GNN}} \\ \hline
      
    CSRM~\cite{wang2019collaborative} & Next item & Memory network & SIGIR 2019 & \scriptsize{\url{https://github.com/wmeirui/CSRM_SIGIR2019}} \\ \hline

    RepeatNet~\cite{ren2019repeatnet} & Next item & RNN, Attention & AAAI 2019 & \scriptsize{\url{https://github.com/PengjieRen/RepeatNet}} \\ \hline
        
    DGRec~\cite{song2019session} & Next item & GNN & WSDM 2019 & \scriptsize{\url{https://github.com/DeepGraphLearning/RecommenderSystems/tree/master/socialRec}} \\ \hline
    
    FGNN~\cite{qiu2019rethinking} &Next item & GNN  &CIKM 2019 & \scriptsize{\url{https://github.com/RuihongQiu/FGNN}} \\ \hline
     
    TAGNN~\cite{yu2020tagnn} &Next item & GNN  &SIGIR 2020 & \scriptsize{\url{https://github.com/CRIPAC-DIG/TAGNN}} \\ \hline 
     
   LESSR~\cite{chen2020handling} &Next item & GNN  &KDD 2020 & \scriptsize{\url{https://github.com/twchen/lessr}} \\ \hline
     
    MKM-SR~\cite{meng2020incorporating} &Next item & RNN, GNN  &SIGIR 2020 & \scriptsize{\url{https://github.com/ciecus/MKM-SR}} \\ \hline
    \end{tabular} 
 \end{threeparttable} 
\end{table*} 
\end{footnotesize}

\begin{footnotesize}
\vspace{-1em}
\begin{table*}[htb]
  \centering
  \caption{\label{tab:data} Commonly used and publicly accessible real-world datasets for SBRSs}
   \vspace{-0.8em}
    \begin{tabular}{|p{1.2cm}|p{1.5cm}|p{1.6cm}|p{1.8cm}|p{1.4cm}|p{2.3cm}|p{1.6cm}|}  
    \hline
     \makebox[1.2cm][c]{\textbf{Domain}} & \makebox[1.5cm][c]{  \textbf{Dataset}}  &\makebox[1.6cm][c]{\textbf{\# sessions}} & \makebox[1.8cm][c]{\textbf{\# interactions}}& \makebox[1.4cm][c]{\textbf{\# items}} &\makebox[2.3cm][c]{\textbf{Avg. session length}} &\makebox[1.6cm][c]{\textbf{Reference}}   \\ \hline
     
      \multirow{5}{0.8cm}{E-commerce} & RSC 2015\footnotemark[1] &1,375,128  & 5,426,961 &28,582 & 3.95 & \cite{ben2015recsys},\cite{ludewig2018evaluation},\cite{wang2019modeling}
      \\\cline{2-7}
      & Tmall\footnotemark[2] & 1,774,729  &13,418,695&425,348& 7.56 &\cite{ludewig2018evaluation},\cite{wang2019modeling}\\\cline{2-7}
      & Tafeng\footnotemark[3] & 19,538  &144,777&5,263& 7.41 &\cite{wang2018attention},\cite{wang2019modeling} \\\cline{2-7}
      & Diginetica\footnotemark[4] & 780,328 &982,961& 43,097 &5.12 &\cite{wu2019session}\\\cline{2-7}
      & RetailRocket\footnotemark[5] & 59,962 & 212,182 & 31,968 & 3.54 &\cite{ludewig2018evaluation}
      \\ \hline

      \multirow{2}{0.8cm}{News}  
       & CLEF 2017\footnotemark[6] & 1,644,442 & 5,540,486 & 742 & 3.37 &\cite{ludewig2018evaluation}\\\cline{2-7}
       &Globo\footnotemark[7] &1,031,167 & 2,930,849&13,092&2.84&\cite{dang2020study}  \\\cline{2-7}
       &Adressa 16G\footnotemark[8] &2,215& 62,908&6,765&28.4 &\cite{zhang2018deep}
       \\\hline

      \multirow{2}{0.8cm}{Music}  
       &Last.FM\footnotemark[9] &169,576 & 2,887,349 &449,037 &17.03 & \cite{dang2020study} \\\cline{2-7}
       
       &30Music\footnotemark[10] & 31,351,954 & 2,764,474& 210,633& 11& \cite{ludewig2018evaluation},\cite{turrin201530music}\\\cline{2-7}
       
       & NowPlaying\footnotemark[11] & 27,005 & 271,177 & 75,169 & 10.04 &\cite{ludewig2018evaluation}
       \\\hline
       
       \multirow{2}{0.8cm}{POI}  
       & Gowalla\footnotemark[12] & -\footnotemark[0] &   245,157 & 6,871 & - &\cite{feng2015personalized}\\\cline{2-7}
       &Foursquare\footnotemark[13] &- & 155,365&2,675&-&\cite{feng2015personalized}  \\ \hline
       
    \end{tabular} 
    \begin{tablenotes}
        \footnotesize 
         \item[1] $^{0}$Raw POI data does not have a session structure, researchers often manually build sessions by treating a user’s check-ins in a single day as a session~\cite{guo2019streaming}.
 \end{tablenotes}
 \vspace{-1.5em}   
\end{table*} 
\end{footnotesize}

\footnotetext[1]{\url{https://www.kaggle.com/chadgostopp/recsys-challenge-2015}}

\footnotetext[2]{\url{https://tianchi.aliyun.com/dataset/dataDetail?dataId=42}}

\footnotetext[3]{\url{https://www.kaggle.com/chiranjivdas09/ta-feng-grocery-dataset}}

\footnotetext[4]{\url{https://competitions.codalab.org/competitions/11161}}

\footnotetext[5]{\url{https://www.kaggle.com/retailrocket/ecommerce-dataset}}

\footnotetext[6]{\url{https://www.newsreelchallenge.org/dataset/}}

\footnotetext[7]{\url{https://www.kaggle.com/gspmoreira/news-portal-user-interactions-by-globocom}}

\footnotetext[8]{\url{http://reclab.idi.ntnu.no/dataset/}}

\footnotetext[9]{\url{http://millionsongdataset.com/lastfm/}}

\footnotetext[10]{\url{http://recsys.deib.polimi.it/datasets/}}

\footnotetext[11]{\url{https://www.kaggle.com/chelseapower/nowplayingrs}}

\footnotetext[12]{\url{http://snap.stanford.edu/data/loc-gowalla.html}}

\footnotetext[13]{\url{https://www.kaggle.com/chetanism/foursquare-nyc-and-tokyo-checkin-dataset}}

Datasets are necessary for evaluating SBRS algorithms. We summarize a collection of 13 publicly accessible real-world datasets that are commonly used for SBRS evaluations in Table~\ref{tab:data} in the supplemental material. These datasets cover a wide range of application domains from e-commerce to POI and are with various characteristics. They can provide challenging test-beds for SBRS algorithms. The collection of algorithms and datasets will be updated on the associated github page~\footnote{\url{https://github.com/shoujin88/SBRS-Survey}}.

\section{Prospects and Future Directions}\label{prospect}
Our comprehensive review of the literature has revealed the significant challenges facing SBRS research and the enormous opportunities SBRS presents. In this section, we outline several promising prospective research directions, which we believe are critical to the further development of the filed.

\subsection{Session-based Recommendations with General User Preference}
$Significance.$ SBRSs usually ignore users' long-term general preferences which can be well captured by conventional RSs like collaborative-filtering based RSs. This may lead to unreliable recommendations since users with different general preferences and consumption habits may chose different items even under the same session context. In this case, how to effectively incorporate users' general preferences into an SBRS is critical yet challenging.

$Open$ $issues.$ Here, we discuss two major issues w.r.t. the general preference learning as well as its incorporation into SBRSs and sketch several critical future research directions.

- How to incorporate users' explicit general preference into an SBRS? In this case, it is assumed that the explicit user-item preference data, e.g., a user-item rating matrix, is available. An intuitive way is to first learn users' general preferences from the explicit preference data using conventional RS approaches, e.g., Matrix Factorization (MF), and then take the learned preference as an indicator to fine-tune the ranking of candidate items in an SBRS, e.g., to put those items more preferred by a user to the front of the recommendation list. Another way is to combine both users' long-term general preferences and short-term preferences together when ranking the candidate items. For instance, a Generative Adversarial Network (GAN) framework was proposed by Zhao et al. \cite{zhao2017leveraging} to
build a hybrid model for movie recommendations. In this model, MF and RNN are utilized to learn users' long-term preference and short-term preference respectively. However, the efforts to address this issue are still limited and more efforts are needed.  

- How to incorporate a user's implicit general preference into an SBRS? In the real word, the explicit preference data may not be always available since users may or may not provide explicit feedback, e.g., ratings, on everything they bought. In such a case, the implicit preference data, i.e., users' transaction behaviour data including view, click, add to cart and purchase, can be leveraged to learn users' implicit general preference \cite{he2016vbpr,he2016fast}. In practice, such implicit preference data is often available in the session-based recommdation scenario \cite{schnabel2018short}. Although a variety of works \cite{anyosa2018incremental,peska2017using} have explored how to learn a users' general preference from such implicit preference data in conventional RSs, e.g., collaborative filtering,
the efforts in SBRS field are still limited. Therefore, how to simultaneously learn a user's implicit general preference and her short-term preference and effecetively integrate them for accurate SBR is a challenging issue which requires more efforts.

\subsection{Session-based Recommendations Considering More Contextual Factors} 
$Significance.$ A context refers to the specific internal and external environment when a user makes choices on items \cite{cao2016non}. Accordingly, contextual factors refer to context-related aspects that may affect a user's choices, such as weather, season, location, time and recent popularity trend. Taking these contextual factors into account may make a huge difference on recommendation performance~\cite{jannach2020research}, as demonstrated by researchers including Adomavicius et al. \cite{adomavicius2015context} and Pagano et al. \cite{pagano2016contextual}. In practice, an SBRS can be seen as a simplified context-aware RS whose context is simplified to a session context \cite{twardowski2016modelling}. Although contextual information has been incorporated into other types of RSs including context-aware RSs \cite{unger2015latent,adomavicius2015context}, most of the contextual factors are rarely exploited in SBRSs.

$Open$ $issues.$ How to incorporate more contextual factors into SBRSs? A few works have provided some initial solutions to this issue. A Contextual RNN for Recommendation (CRNN) was proposed to incorporate contextual factors including the time gap between different interactions and the time of a day when an interaction happens, into an RNN-based SBRS \cite{smirnova2017contextual}. In several works \cite{lerche2016value,jannach2017determining}, the recent popularity trend, a user's recently viewed items, and items with discount in shopping mall are taken as contextual factors for SBR. However, these solutions are just a starting point, and more explorations are still necessary to address issues such as how to collect more contextual information and how to more effectively incorporate it for more accurate SBR.

\subsection{Session-based Recommendations with Cross-domain Information}\label{domainRS}
$Significance.$ Cross domains mean different but relevant domains \cite{hu2013cross,zhu2021cross}, e.g., movie domain and song domain. Usually, a user's purchased items come from multiple domains rather than a single one to meet her demand~\shortcite{zhu2019dtcdr}. In addition, a user's choices on items from different domains are often dependent. For example, after Alice watches the movie ``Titanic", she may listen to the movie's theme song ``My heart will go on''. Such an example shows that items from different domains may not only be dependent but can even form a sequential session, such as \{``Titanic", ``My heart will go on''\}. The recommendations based on such type of session data are interesting but quite challenging. Such recommendations not only cover more aspects of our daily lives but also provide a solution to the data sparsity issue when only one domain is considered. On the other hand, it is hard to collect a user's consumed items from various domains together, and also the relations between items from different domains are much more complex than that from a single domain. 

$Open$ $issues.$  According to whether the items from different domains can form a session or not, there are two main open issues to be explored.

- How to borrow knowledge from other domains to benefit the SBR in the target domain? When no sessions can be built across different domains, a target-auxiliary framework can be employed to benefit SBR from other domains. To be specific, the framework takes the target domain where the recommendations are made as the main information source while taking other domains as a supplementary one. An intuitive instance of the framework could utilize transfer learning \cite{pan2010survey,elkahky2015multi} which transfers knowledge from source domains to help with the tasks in the target one. Although transfer learning has been well explored in conventional RSs like collaborative filtering \cite{pan2013transfer,loni2014cross}, it is rarely explored in SBRSs. 

- How to perform SBR on multiple domains? This case happens when sessions can be built from different domains. Different from the aforementioned target-auxiliary framework, in this case, RSs treat items from different domains equally and each domain can serve as the target domain to make recommendations. This is much more interesting yet challenging than the first case. Consequently, how to develop advanced models to effectively capture the complex and heterogeneous dependencies \cite{cao2015coupling} among different domains for accurate SBR requires more explorations.  

\subsection{Session-based Recommendations by Considering More User Behaviour Patterns}
$Significance.$ In addition to the basic co-occurrence or sequential behaviour patterns hidden in session data, there are actually more types of user behaviour patterns that can be leveraged to benefit SBR, such as repeat consumption \cite{ren2019repeatnet} and periodic consumption \cite{hu2020modeling}. Such kinds of behaviour patterns are not uncommon in the real world, but are usually ignored by existing works on SBRSs.        

$Open$ $issues.$ How to effectively discover and leverage more types of user behaviour patterns to improve SBR? On one hand, the other types of patterns are less frequent and obvious than the basic co-occurrence or sequential patterns which are thus more difficult to be precisely identified. On the other hand, how these patterns can influence the users' final choices on items is not very clear. Hence, it is difficult to effectively incorporate the useful information from these patterns while reducing noise to benefit the recommendations. Such issues are of practical significance in real-world applications, especially in the e-commence industry, and thus require more efforts.

\subsection{Session-based Recommendations with Constraints }
$Significance.$ In reality, it is not uncommon that there are some types of underlying constraints over the items in one session. For example, in some cases, the items purchased in one session may not be identical or similar, instead, they may complement each other \cite{zhao2018deep}, e.g., milk and bread, to form a coherent package to satisfy a user's certain goal, e.g., breakfast. In other cases, there may be duplicated items within a session, since a user may buy multiple copies of items in one session. Such kinds of constraints are often ignored by existing works on SBRSs. 

$Open$ $issues.$ How to generate a session with some constraints on it to better satisfy a user's purchase goal? On one hand, different users may have different interaction patterns and requirements to be satisfied with different constraints. Therefore, it is difficult to determine which constraints should be added to a specific user to best accomplish her goal. On the other hand, it is also challenging to jointly optimize both the added constraints and the prediction accuracy. This open issue is challenging but of practical concern, with only limited attention from the community~\cite{quadrana2018sequence}. A straightforward solution is to incorporate some semantic relations between items by employing a knowledge graph \cite{wang2018ripplenet} into SBRSs.

\subsection{Interactive Session-based Recommendations}
$Significance.$ In the real-word cases, particularly in the online shopping scenario, a session is often generated from a continuous interactive process between a user and the shopping platform. For example, a user may first click an item to start a session, followed by different actions on this item, e.g., view it, add it to cart, or just skip it. By taking these different actions as different feedback from the user, the platform can then accordingly adjust recommendation strategies for the subsequent items. Such interactive process proceeds till to the end of the session. Though such intrinsic nature embedded in session generation process is quite important for precisely learning the user's dynamic preference for accurate SBR, it has been overlooked by most of the existing works on SBRSs.      

$Open$ $issues.$ How to effectively model the continuous interactive process between users and the platform for interactive SBR? This issue is critical but challenging in the area of SBRSs. In such a case, a user's preference is revealed by her limited real-time interactions with the platforms and usually changes over time. So, it is difficult to precisely capture the user's dynamic preferences from limited interactions in a timely manner. Although, some researchers have proposed approaches based on reinforcement learning \cite{zou2020pseudo} to address it, the studies are still in early stage while more and deeper explorations are needed.

\subsection{Online or Streaming Session-based Recommendations }
$Significance.$ On a real-word online shopping platform, session data usually comes incrementally in a streaming scenario. This leads to the continuous, large-volume, high-velocity nature of session data \cite{guo2019streaming,zhao2020double}. However, most of the existing studies on SBRSs work on the offline and static data, which may be inconsistent with the real-world application scenario.       

$Open$ $issues.$ How to effectively learn users' dynamic preferences in an online and streaming scenario for better SBR? This is much more challenging than the SBRSs based on offline data, but of greater significance from the practical perspective. Specifically, it is quite challenging to develop accurate and highly-efficient recommendation algorithms to effectively model large-volume streaming data and generate timely recommendations. Researchers tried to address this issue with adaptively distilled exemplar replay strategy~\cite{mi2020ader}, a continuously queried and updated non-parametric
memory mechanism \cite{mi2020memory} or a reservoir-based
streaming model \cite{guo2019streaming,qiu2020gag}. More efforts are still required for this open issue.

\section{Conclusions}\label{Conclusions}
In this paper, we have conducted a systematic and extensive review of the most notable works to date on session-based recommender systems (SBRSs). We have proposed a unified framework to organise the existing works in this area into three sub-areas and provided a unified problem statement for SBRS to reduce some confusions and inconsistencies in the field. We have thoroughly analyzed the characteristics of session data and the corresponding challenges they bring for SBRSs. We have also proposed a classification scheme for the organization and clustering of existing approaches for SBRSs, and highlighted some critical technical details for each class of approaches. In addition, we have discussed some of the most pressing open issues and promising directions. The research in SBRS field is flourishing and a number of newly developed techniques and emerging approaches keep coming. It is our hope that this survey can provide readers with a comprehensive understanding of the key aspects, main challenges, notable progress in this area, and shed some light on future studies.




\begin{acks}
The authors would like to thank Dr. Liang Hu and Mr. Yan Zhao for their constructive suggestions on this work. This work was supported by Australian Research Council
Discovery Grants (DP180102378, DP190101079 and FT190100734).
\end{acks}

\normalem
\small{
\bibliographystyle{ACM-Reference-Format}
\bibliography{Reference}
}